\documentclass[fleqn,usenatbib]{mnras}

\usepackage{newtxtext,newtxmath}

\usepackage[T1]{fontenc}
\usepackage{ae,aecompl}

\usepackage{graphicx}	
\usepackage{amsmath}	
\usepackage{float}
\usepackage{natbib}
\usepackage{color,soul}
\usepackage{adjustbox,lipsum}
\usepackage{array}
\usepackage{cleveref}
\usepackage{booktabs}
\usepackage[flushleft]{threeparttable}
\usepackage{caption}

\title[AGN and SF: Data versus Theoretical Models]{AGN and Star Formation at Cosmic Noon: Comparison of Data to Theoretical Models}

\author[Florez et al.]{
Jonathan Florez,$^{1}$\thanks{E-mail: jflorez06@utexas.edu}
Shardha Jogee,$^{1}$ 
Yuchen Guo,$^{1}$ 
Sofía A. Cora,$^{2,3}$
\newauthor
Rainer Weinberger,$^{4}$
Romeel Davé,$^{5}$
Lars Hernquist,$^{4}$
Mark Vogelsberger,$^{6}$
\newauthor
Robin Ciardullo,$^{7,8}$
Steven L. Finkelstein,$^{1}$
Caryl Gronwall,$^{7,8}$
\newauthor
Lalitwadee Kawinwanichakij,$^{9}$
Gene C.K. Leung,$^{1}$
Stephanie LaMassa,$^{10}$
\newauthor
Casey Papovich,$^{11}$ 
Matthew L. Stevans,$^{1}$ 
and Isak Wold$^{12}$
\\
$^{1}$Department of Astronomy, University of Texas at Austin, Austin, TX 78712, USA\\
$^{2}$Instituto de Astrofísica de La Plata (CCT La Plata, CONICET, UNLP), Observatorio Astronómico, Paseo del Bosque, B1900FWA La Plata, Argentina\\
$^{3}$Facultad de Ciencias Astronómicas y Geofísicas, Universidad Nacional de La Plata, Observatorio Astronómico, Paseo del Bosque, B1900FWA La Plata, Argentina\\
$^{4}$Center for Astrophysics -- Harvard \& Smithsonian, 60 Garden Street, Cambridge, MA 02138, USA\\
$^{5}$Institute for Astronomy, Royal Observatory, University of Edinburgh, Edinburgh EH9 3HJ, UK\\
$^{6}$Department of Physics, Kavli Institute for Astrophysics and Space Research, MIT, Cambridge, MA 02139, USA\\
$^{7}$Department of Astronomy and Astrophysics, The Pennsylvania State University, University Park, PA 16802, USA\\
$^{8}$The Institute for Gravitation and the Cosmos, The Pennsylvania State University, University Park, PA 16802, USA\\
$^{9}$Kavli Institute for the Physics and Mathematics of the Universe, The University of Tokyo, Kashiwa, Japan 277-8583 (Kavli IPMU, WPI)\\
$^{10}$Space Telescope Science Institute, 3700 San Martin Dr, Baltimore, MD 21218, USA\\
$^{11}$Department of Physics and Astronomy, Texas A\&M University, College Station, TX 77843, USA\\
$^{12}$NASA Goddard Space Flight Center, Greenbelt, MD 20771
}

\date{Accepted XXX. Received YYY; in original form ZZZ}

\pubyear{2021}

\begin{document}

\defcitealias{2020MNRAS.497.3273F}{Paper 1}
\defcitealias{2007ApJ...654..731H}{H07}
\defcitealias{2012MNRAS.425..623L}{L12}

\label{firstpage}
\pagerange{\pageref{firstpage}--\pageref{lastpage}}
\maketitle


\begin{abstract}
In theoretical models of galaxy evolution, AGN and star formation (SF) activity are closely linked and AGN feedback is routinely invoked to regulate galaxy growth. In order to constrain such models, we compare the hydrodynamical simulations IllustrisTNG and SIMBA, and the semi-analytical model SAG to the empirical results on AGN and SF at cosmic noon ($0.75 < z < 2.25$) reported in \cite{2020MNRAS.497.3273F}. The empirical results are based on a large mass-complete sample drawn from 93,307 galaxies with and without high X-ray luminosity AGN ($L_X \gtrsim 10^{44}$ erg s$^{-1}$), selected from a 11.8 deg$^2$ area ($\sim 0.18$ Gpc$^3$ comoving volume at $z=0.75-2.25$). The main results of our comparisons are: (i) SAG and IllustrisTNG both qualitatively reproduce the empirical result that galaxies with high X-ray luminosity AGN have higher mean SFR, at a given stellar mass, than galaxies without such AGN. SAG, however, strongly over-produces the number density of high X-ray luminosity AGN by a factor of 10 to 100, while IllustrisTNG shows a lack of high X-ray luminosity AGN at high stellar mass ($M* > 10^{11} \ M_{\odot}$) at $z \sim 2$. (ii) In SIMBA, the mean SFR of galaxies with high X-ray luminosity AGN is lower than the SFR of galaxies without such AGN. Contrary to the data, many high X-ray luminosity AGN in SIMBA have quenched SF, suggesting that AGN feedback, or other feedback modes in galaxies with such AGN, might be too efficient in SIMBA.
\end{abstract}




\section{Introduction} \label{introduction}

Understanding the connection between galaxy properties and their central massive black holes (BHs) currently remains one of the biggest challenges to formulating theoretical models of galaxy evolution. Observations have shown that the central BH mass of a galaxy correlates with the galaxy bulge stellar mass \citep{1998AJ....115.2285M, 2002MNRAS.331..795M} and galaxy bulge velocity dispersion \citep{2000ApJ...539L...9F, 2000ApJ...539L..13G, 2013ARA&A..51..511K}, suggesting a clear link between galaxy and BH growth. Furthermore, the cosmic black hole accretion rate (BHAR) density and star formation rate (SFR) density are observed to peak at $z \sim 2$ and decline in tandem down to $z \sim 0$ \citep{2007A&A...474..755B, 2008MNRAS.385..687W, 2009ApJ...697.1971J, 2014ARA&A..52..415M, 2014MNRAS.439.2736D}. Such trends suggest that galaxy and BH growth are closely intertwined, but whether they coevolve remains a topic of debate \citep[e.g.,][]{2013ARA&A..51..511K, 2011ApJ...734...92J}.

Active galactic nuclei (AGN) activity in galaxies arises directly from the accretion of gas onto a galaxy's central massive BH. The resulting feedback from an AGN is thought to suppress or reduce star formation (SF) in the host galaxy via jets, winds, and radiation. In theoretical simulations of galaxy evolution, AGN feedback is often invoked, along with stellar feedback, in order to solve the ``overcooling'' problem in galaxy formation models whereby galaxies grow too massive and produce stellar mass functions that do not resemble empirical ones unless some form of heating or feedback is applied to the cooling gas inside the dark matter halo \citep{2017ARA&A..55...59N, 2015ARA&A..53...51S}. Different forms of AGN feedback are postulated to affect the host galaxy and the gas in the surrounding environment in different ways. Winds and radiation from the accretion disk near the BH can heat gas or expel it on various galactic scales \citep{2016MNRAS.458..816H, 2015ApJ...800...19R, 2015MNRAS.449.4105C, 2013MNRAS.436.3031V, 2012ARA&A..50..455F, 2011ApJ...738...16H} whereas the jets from an AGN can heat the gas in the intracluster medium (ICM) of galaxy clusters, therefore preventing gas cooling and ultimately preventing future episodes of star formation \citep{2019MNRAS.486.2827D, 2014ARA&A..52..589H, 2012ARA&A..50..455F, 2006PhR...427....1P}. Cosmological simulations often try to model these two forms of AGN feedback (i.e., quasar mode and radio/jet mode), however, lack of high spatial and spectral resolution observations of AGN along with poor simulation resolution has made it difficult to constrain the physical mechanisms driving AGN feedback, BH growth, and the extent to which AGN feedback can heat and/or expel gas. It should be noted, however, that some numerical simulations do model AGN feedback parameters based on observations of outflows of molecular and ionised gas \citep[e.g.,][]{2017A&A...603A..99P, 2013MNRAS.436.2576L, 2011ApJ...733L..16S} as well as observations of radio-loud jets \citep{2012ARA&A..50..455F} in galaxies with bright AGN \citep{2019MNRAS.486.2827D}.

While it has been postulated that some stages of AGN activity might suppress SF as outlined above, there is evidence of AGN and enhanced SF activity coexisting in galaxies. Many studies have found, at low and high redshift, luminous AGN residing in galaxies with enhanced episodes of SF \citep{2020MNRAS.497.3273F, 2018A&A...618A..31M, 2017MNRAS.471.3226M, 2013ApJ...771...63R, 2012A&A...540A.109S, 1988ApJ...325...74S}. Other studies, however, provide evidence for decreased SFRs in galaxies with luminous AGN \citep{2016MNRAS.455L..82L, 2015MNRAS.452.1841S}. An enhancement of SF activity in AGN host galaxies could at least be, in part, due to processes that are capable of igniting and fueling circumnuclear SF and BH accretion when the angular momentum problem can be overcome \citep[e.g.,][and references therein]{2006LNP...693..143J}. Examples of such processes include gas-rich major and minor mergers \citep{2008ApJS..175..356H}, gravitational torques from bars and/or non-axisymmetric features in the disc, stellar feedback, and hydromagnetic winds \citep[see][and references therein]{2006LNP...693..143J}. It is critical for models of galaxy evolution to constrain how AGN feedback impacts galaxy formation, especially since AGN and SF are always assumed to be closely linked in such models and AGN feedback is routinely invoked to regulate galaxy growth.

Until recently, analyzing the SF properties of galaxies with bright AGN proved rather difficult as AGN emission is capable of dominating the galaxy spectral energy distribution (SED) across a wide range of wavelengths (e.g., UV to mid-IR), thereby making it difficult to distinguish between AGN activity and emission from SF in the SED fitting. For this reason, studies of galaxy evolution often remove galaxies with bright AGN from their samples \citep{2021arXiv210314690S, 2021MNRAS.505..947S, 2020ApJ...892....7K, 2020MNRAS.499.4239S, 2020MNRAS.491.3318S, 2013A&A...556A..55I} or fail to analyze galaxies with and without high luminosity AGN using a consistent methodology or a statistically large enough sample \citep{2014ApJ...795..104W, 2013ApJ...777...18M, 2012ApJ...754L..29W, 2011A&A...533A.119E, 2009ApJ...691.1879W, 2007A&A...468...33E, 2007ApJS..173..267S,2004MNRAS.351.1151B}. In recent years, however, numerous SED fitting codes have been developed that are capable of disentangling the emission from AGN and SF in the SED and producing more reliable estimates of stellar mass and SFR (e.g., CIGALE \citep{2019A&A...622A.103B}, AGNfitter \citep{2016ApJ...833...98C}, SED3FIT \citep{2013A&A...551A.100B}). In \cite{2020MNRAS.497.3273F}, we used CIGALE to perform SED fitting on a large sample of massive galaxies with and without high X-ray luminosity AGN at $z = 0.5-3$. By using CIGALE we were able to derive galaxy properties, such as SFR and stellar mass, for galaxies with and without high X-ray luminosity AGN in a self-consistent manner. 

The goal of this project is to follow up on the work done in \cite{2020MNRAS.497.3273F} (hereafter \citetalias{2020MNRAS.497.3273F}), which reports two key results: (i.) At fixed stellar mass, galaxies with high X-ray luminosity AGN ($L_X > 10^{44}$ erg s$^{-1}$) have a mean SFR that is a factor of $\sim 3-10$ times higher than galaxies without such AGN; (ii.) The majority ($> 95\%$) of AGN with high X-ray luminosity  do not reside in galaxies with quenched star formation. These results suggest that enhanced SFRs and AGN activity in galaxies are triggered by mechanisms that lead to large gas inflow rates on kpc to sub-pc scales (e.g., mergers), growing both the supermassive BHs (SMBHs) and the stellar masses of the host galaxies. These results also suggest that if AGN feedback quenches the SF in a galaxy, it does so after the high X-ray luminosity phase of AGN activity. In this paper, we will expand on these results by analyzing the cosmological hydrodynamical simulations IllustrisTNG \citep{2018MNRAS.475..624N, 2018MNRAS.475..676S, 2018MNRAS.477.1206N, 2018MNRAS.475..648P, 2018MNRAS.480.5113M, 2019ComAC...6....2N, 2019MNRAS.490.3234N, 2019MNRAS.490.3196P} and SIMBA \citep{2019MNRAS.486.2827D}, as well as the semi-analytical model of galaxy formation and evolution SAG \citep{2018MNRAS.479....2C}. We explore the following questions: (i) How does the number density of high X-ray luminosity AGN in theoretical models and numerical simulations compare to the empirical number density of observed AGN (Section \ref{xlf})? (ii) Do theoretical models reproduce the observed distribution of galaxies with and without high X-ray luminosity AGN in the stellar mass–SFR plane as well as the empirical result that galaxies with high X-ray luminosity AGN have a higher SFR, on average, at fixed stellar mass than galaxies without such AGN (Section \ref{mass_sfr})? 

This paper is organized as follows: In Section \ref{data}, we summarize the results of \citetalias{2020MNRAS.497.3273F} and briefly outline the data, sample selection, SED fitting procedure, and stellar mass and SFR completeness limit estimates used in that paper. In Section \ref{sample_updates}, we discuss the updates to the sample selection in \citetalias{2020MNRAS.497.3273F} and in Section \ref{xray_dems} we discuss the observed demographics of the high X-ray luminosity AGN population. In Section \ref{simulation_data} we discuss the three theoretical models, their implementation of AGN and SF feedback, and how we obtain the X-ray luminosity from the black hole accretion rate. In Section \ref{results} we present our results, and in Sections \ref{discussion} and \ref{summary} we discuss and summarize our results, respectively. In this paper we assume $H_0 = 70$ km s$^{-1}$ Mpc$^{-1}$, $\Omega_{\rm M} = 0.3$, and $\Omega_{\Lambda} = 0.7$ for the observed data \citepalias[as in][]{2020MNRAS.497.3273F}.

\section{Overview of Empirical Results and Data Analysis in Paper 1} \label{data}
The goal of the present paper is to evaluate the extent to which three state-of-the-art theoretical models can reproduce the empirical results presented in \citetalias{2020MNRAS.497.3273F} \citep{2020MNRAS.497.3273F}. In this section, we summarize the empirical results of \citetalias{2020MNRAS.497.3273F} and outline the associated data, sample selection, and methodology we follow for the analysis. In \citetalias{2020MNRAS.497.3273F}, we compared the stellar masses and SFRs of a large mass-complete sample of galaxies with and without high X-ray luminosity AGN at $0.5 < z < 3$ (258 galaxies complete in X-ray luminosity with $L_X \gtrsim 10^{44}$ erg s$^{-1}$, 153,765 galaxies without). Our data were selected from a relatively large $\sim 11.8$ deg$^2$ area in Stripe 82 with multiwavelength data (X-ray to far-IR) available, corresponding to a large comoving volume of $\sim 0.3$ Gpc$^3$ at $0.5 < z < 3$. A large strength of this work was that we self consistently modeled and fit the SEDs of galaxies with and without high X-ray luminosity AGN using the SED fitting code CIGALE. As mentioned in the introduction, two of the key results we reported in \citetalias{2020MNRAS.497.3273F} are: (i.) The mean SFR, at fixed stellar mass, of galaxies with high X-ray luminosity AGN is higher by a factor of $\sim 3-10$ than the mean SFR of galaxies without such AGN; (ii.) An overwhelming number of galaxies with high X-ray luminosity AGN ($> 95 \%$) do not have quenched SF.

The data, sample selection, SED fitting, and stellar mass and SFR completeness limit estimates used in the present paper are essentially the same as those used in \citetalias{2020MNRAS.497.3273F}. We give a brief overview of these topics in the following subsections. Before comparing the empirical results of \citetalias{2020MNRAS.497.3273F} to the theoretical models, we make some small adjustments to the sample selection in order to improve the robustness of these comparisons. These adjustments will be discussed in Section \ref{sample_updates}.

\subsection{Photometric Catalogs}
Our photometric data are primarily obtained from the NEWFIRM $K_S$-selected catalog ($5 \sigma$ depth of 22.4 AB mag) of \cite{2021arXiv210314690S} which covers 17.5 deg$^2$ of the Sloan Digital Sky Survey (SDSS) Stripe 82 equatorial field. The NEWFIRM $K_S$-selected catalog includes $u,g,r,i,z$ photometry from DECam as well as 3.6 and 4.5 $\mu$m photometry from IRAC. We supplement this photometry with $J$ and $K_S$ data from VICS82 in order to obtain a total of ten photometric filters spanning near-UV to mid-IR wavelengths. In \citetalias{2020MNRAS.497.3273F} we ran the EAZY-py SED fitting code \citep{2008ApJ...686.1503B} on this data to obtain photometric redshifts for galaxies without a high X-ray luminosity AGN. The photometric redshifts have an accuracy of $\sigma_z = 0.037$ at $z < 1$ \citepalias{2020MNRAS.497.3273F}, and a photometric redshift accuracy of $\sigma_z =  0.102$ at $1.9 < z < 3.5$ \citep{2021MNRAS.505..947S}.

We utilize X-ray photometry from the Stripe 82X X-ray survey \citep{2016ApJ...817..172L} in order to identify and analyze galaxies with high X-ray luminosity AGN. Stripe 82X covers 31.3 deg$^2$ of the SDSS Stripe 82 equatorial field and includes archival $Chandra$ and XMM-$Newton$ X-ray data. For this project, we use photometry from the XMM-$Newton$ Announcement Opportunity 13 (AO13) that was introduced in \cite{2016ApJ...817..172L}. We crossmatch the Stripe 82X AO13 data to the NEWFIRM $K_S$-selected catalog as described in \citetalias{2020MNRAS.497.3273F} using the maximum likelihood estimator (MLE) method of \cite{1992MNRAS.259..413S}. Spectroscopic redshifts for this sample are obtained from \cite{2019ApJ...876...50L} with $\sim 70 \%$ completeness. The remaining $\sim 30 \%$ of sources without a spectroscopic redshift have a photometric redshift obtained from \cite{2017ApJ...850...66A}. The photometric redshifts presented in \cite{2017ApJ...850...66A}, when compared to available spectroscopic redshifts, have a normalized median absolute deviation of $\sigma_{\rm NMAD} \sim 0.06$ at $0 < z < 3$. 

\subsection{Sample Selection} \label{sample_selection}
For the analysis in \citetalias{2020MNRAS.497.3273F} we produced two final samples of galaxies: a sample of galaxies with high X-ray luminosity AGN, and a control sample of galaxies without such AGN. Both samples were chosen from the region of the sky where the $Spitzer$-HETDEX Exploratory Large Area \citep[SHELA,][]{2016ApJS..224...28P} and the Stripe 82X footprints overlap, spanning an area of 11.8 deg$^2$, corresponding to a comoving volume $\sim 0.3$ Gpc$^3$ at our redshift range of interest $0.5 < z < 3$. In \citetalias{2020MNRAS.497.3273F} we did not correct our sample of galaxies with high X-ray luminosity AGN for dust obscuration. We cited \cite{2020ApJ...891...41P} who calculated how many AGN would be added to their high X-ray luminosity ($L_X > 10^{44.5}$ erg s$^{-1}$) sample at $z > 1$ if they made a correction for dust obscuration and found that their sample would increase by very little ($\lesssim 4\%$). We therefore believe our sample of high X-ray luminosity AGN would similarly not increase by much if we corrected for dust obscuration.

\begin{table*}
\begin{center}
\begin{threeparttable}
\caption{The number of galaxies with and without high X-ray luminosity AGN at each redshift range.}
\begin{tabular}{l c c c c}
\hline
\hline
Sample & All $z$ bins ($0.75 < z < 2.25$) & $0.75 < z < 1.25$ & $1.25 < z < 1.75$ & $1.75 < z < 2.25$ \\
(a) & (b) & (c) & (d) & (e)\\
\hline
\hline
(1) S0-DECam-NEWFIRM-IRAC & 209,721 & 130,650 & 53,074 & 25,997\\
\hline
(2) S1-Lum-AGN\\
i.) Total Number $N_{\rm{tot}}$ & 667 & 321 & 228 & 118\\
ii.) $L_X > L_{X,\rm{lim}}$ & 321 & 151 & 117 & 54\\
iii.) $L_X > L_{X,\rm{lim}}$ \& $M_* > M_{*,95\% \rm{lim}}$ & 203 & 119 & 66 & 18\\
\hline
(4) S2-No-Lum-AGN \\
i.) Total Number $N_{\rm{tot}}$ & 209,054 & 130,329 & 52,846 & 25,879\\
ii.) $M_* > M_{*,95\% \rm{lim}}$ & 93,104 & 67,015 & 21,330 & 4,759\\
\hline
\end{tabular}
\begin{tablenotes}
\small
\item *Note: (1) S0-DECam-NEWFIRM-IRAC contains all galaxies in SHELA that fall inside the Stripe 82X survey footprint and have a detection in the DECam $u,g,r,i,z$ bands, a $\rm S/N > 5$ in the NEWFIRM K-band, a $\rm S/N > 2$ in the 3.6 and 4.5 $\mu$m IRAC bands. (2) S1-Lum-AGN contains the subsample of galaxies in S0-DECam-NEWFIRM-IRAC that have a high X-ray luminosity AGN{}. The total number of galaxies and the number of galaxies with X-ray luminosities above the completeness limit ($L_X > L_{X,\rm{lim}}$) and stellar masses above the $95 \%$ stellar mass completeness limit ($M_* >M_{*,95 \%\rm{lim}}$) are also shown. (3) S2-No-Lum-AGN contains the subsample of galaxies in S0-DECam-NEWFIRM-IRAC that do not have a high X-ray luminosity AGN{}.
\end{tablenotes}
\label{tab1}
\end{threeparttable}
\end{center}
\end{table*}

For the final samples we used in our analysis in \citetalias{2020MNRAS.497.3273F}, we required a $K_S$ flux with a signal-to-noise ratio (S/N) detection greater than 5, an IRAC 3.6 and 4.5 $\mu$m flux with a S/N detection greater than 2, and a detection in the DECam $u,g,r,i,z$ filters. We required this for both samples of galaxies with and without high X-ray luminosity AGN. These sample cuts gave us a total of 932 galaxies with high X-ray luminosity AGN at $z=0.5-3$ and a total of 318,904 galaxies without high X-ray luminosity AGN at the same redshift range. We clarify here that the sample of galaxies without high X-ray luminosity AGN that we analyze in \citetalias{2020MNRAS.497.3273F} does not include any galaxies with X-ray emission detected in XMM AO13 (see Figure 1 in \citetalias{2020MNRAS.497.3273F}). This sample, however, may include galaxies hosting AGN whose X-ray luminosity are lower and lie below the X-ray detection limit of XMM AO13.

\subsection{SED Fitting}
One of the large advantages of \citetalias{2020MNRAS.497.3273F} is that we fit the SEDs of galaxies with and without high X-ray luminosity AGN using the exact same SED fitting code. This allowed us to self-consistently derive and obtain stellar masses and SFRs for galaxies with and without high X-ray luminosity AGN. In our SED fitting of galaxies with high X-ray luminosity AGN, we implemented the \cite{2006MNRAS.366..767F} AGN emission templates to model the AGN emission at UV to mid-IR wavelengths. We did not include AGN emission templates in the SED fitting of galaxies without high X-ray luminosity AGN, however. For all galaxies, we included models of dust emission due to star formation from \cite{2014ApJ...784...83D} and the stellar population synthesis models of \cite{2003MNRAS.344.1000B}. We assumed dust attenuation to the SED as described by \cite{2000ApJ...533..682C}, a Chabrier initial mass function \citep[IMF;][]{2003PASP..115..763C}, and a delayed exponential SFH \citep{2018A&A...615A..61C,2016A&A...585A..43C}.

In \citetalias{2020MNRAS.497.3273F}, we performed a number of tests to determine if our CIGALE-derived SFRs were reliable. First, we ran CIGALE on a sample of 38 galaxies with high X-ray luminosity AGN from the CANDELS survey \citep{2011ApJS..197...35G, 2011ApJS..197...36K}, using photometric bandpasses that span the same wavelength range as our own sample, and compared the CIGALE-derived SFRs to the SFRs from \cite{2019MNRAS.485.3721Y, 2017ApJ...842...72Y} for the same group of galaxies. We find relatively good agreement between our CIGALE-derived SFRs and those presented in \cite{2019MNRAS.485.3721Y, 2017ApJ...842...72Y}, who obtain SFRs by taking the median value of the SFR measured by numerous teams for the same object (see Section 4.3 of \citetalias{2020MNRAS.497.3273F}). Second, we created mock galaxy SEDs with AGN emission included and then ran CIGALE on the mock galaxy SED fluxes. We then compared the CIGALE-derived SFRs to the true mock galaxy SFRs and found that $\sim 90 \%$ of the CIGALE-derived SFRs agree with the true mock galaxy SFRs to within a factor of 3-4 and only $2 \%$ of the CIGALE-derived SFRs differ from the true SFRs by a factor 10 or more (see Section 4.4 of \citetalias{2020MNRAS.497.3273F}). Lastly, we ran EAZY-py on the sample of galaxies without high X-ray luminosity AGN and compared the EAZY-py SFRs to the CIGALE-derived SFRs. We found that EAZY-py SFRs are consistent with CIGALE SFRs, on average, to within a factor $\sim 2-3$ (see Section 5.2 of \citetalias{2020MNRAS.497.3273F}). These tests demonstrated that our SED fitting procedure in \citetalias{2020MNRAS.497.3273F} is robust and produces reliable SFRs.

\subsection{Estimate of Stellar Mass and SFR Completeness Limits}
In \citetalias{2020MNRAS.497.3273F}, we estimated the stellar mass completeness limits for our sample of galaxies with and without high X-ray luminosity AGN following the procedures described in \cite{2013A&A...558A..23D, 2010A&A...523A..13P}. This method assumes that the stellar mass completeness limit of a galaxy survey can be determined from the least massive galaxy that can be detected in a given bandpass with a magnitude equal to the magnitude limit of the survey in that bandpass. Because our samples were constructed from a $K_S$-selected catalog, we determined the stellar mass completeness limit at each redshift using the $K_S$-band magnitudes of our sources. We refer the reader to \citetalias{2020MNRAS.497.3273F} for a full description of our stellar mass completeness limit estimate. The $95\%$ stellar mass completeness limits for the sample used in this work are $\log(M_*/M_{\odot}) =$ 10.27, 10.58, and 11.03 in our $0.75 < z < 1.25$, $1.25 < z < 1.75$, and $1.75 < z < 2.25$ bins, respectively.

For every galaxy in our sample, we derived a dust-corrected SFR from our SED fitting procedure using all available band-passes. Unlike stellar mass, where there is a fairly direct connection between a galaxy's stellar mass and its $K_S$-band magnitude, it is not so easy to obtain a survey completeness limit for a dust-corrected SFR as this would require the use of many different observed bandpasses and SED models. To overcome this, in \citetalias{2020MNRAS.497.3273F} we used the DECam $u$-band and $g$-band as a proxy for FUV flux and obtained a dust extincted SFR completeness limit based on the $u$-band and $g$-band $5 \sigma$ limiting magnitudes ($m_{u, \rm lim} = 25.0$ AB mag and $m_{g, \rm lim} = 24.8$ AB mag, see \cite{2019ApJS..240....5W}) by applying the SFR$_{\rm FUV}$ conversion factor from \cite{2011ApJ...741..124H}. This conversion factor assumes a \cite{2001MNRAS.322..231K} IMF, so we reduce the estimated SFR$_{\rm FUV}$ by 0.046 dex to make the results consistent with the \cite{2003PASP..115..763C} IMF used throughout this work. For our analysis we used the $u$-band flux at $z < 1.25$ and the $g$-band flux at $z > 1.25$ to estimate the FUV flux. The estimated $u$-band and $g$-band based dust extincted FUV SFR completeness limits are $\log($SFR$_{\rm{FUV, lim}} / M_{\odot}$ yr$^{-1} ) =$ -0.02, 0.33, and 0.62 at $z = 0.75-1.25$, $z=1.25-1.75$ and $z=1.75-2.25$, respectively. We emphasize here that the SED-derived SFRs and the dust-extincted SFRs estimated from the FUV flux are very different quantities. The SED-derived SFRs are likely going to be higher than the dust-extincted SFRs derived from the FUV flux as those from the SED fitting have been corrected for dust extinction. If we apply a dust correction based on the median attenuation measured from CIGALE for galaxies within 0.1 mag of the $5 \sigma$ $u$ and $g$-band completeness limits, which is $\sim 2 - 2.5$ for all three redshift ranges, we find dust-corrected FUV completeness limits of $\log($SFR$_{\rm{FUV,lim}} / M_{\odot}$ yr$^{-1} ) =$ 1.02, 1.23, and 1.40 at $z = 0.75-1.25$, $z=1.25-1.75$ and $z=1.75-2.25$, respectively. We note that we do not use a signal-to-noise ratio cut for any of the DECam filters in our analysis, meaning it would not be unusual to measure intrinsic SFRs from the SED fitting well below the FUV-based SFR completeness limits listed here.

\section{Updates to the Sample Selection} \label{sample_updates}

As mentioned in Section \ref{introduction}, the goal of the present paper is to evaluate the extent to which three state-of-the-art theoretical models can reproduce the empirical results presented in \citetalias{2020MNRAS.497.3273F}. In order to increase the robustness of the comparison between our empirical results and the theoretical models, we make some small updates to the sample selection which we discuss here.

First, we use a narrower redshift range here than the redshift range used in \citetalias{2020MNRAS.497.3273F}, which corresponds to $0.5 < z < 3$. This is because we only have simulation snapshots available at $z=1.0, \ 1.5,$ and 2.0 for the three theoretical models we use in this work, meaning that a more direct comparison of the empirical results to the simulated data can be made at $0.75 < z < 2.25$.  Second, the recent availability of spectroscopy by the Hobby Eberly Telescope Dark Energy Experiment \citep[HETDEX,][]{2008ASPC..399..115H} has allowed us to replace the photometric redshifts of three galaxies from the Stripe 82X sample with spectroscopic redshifts. Lastly, because black hole accretion is intimately connected with the emission produced in the ``hard'' X-ray spectral range of the galaxy SED \citep{1991ApJ...380L..51H}, we select our sample of galaxies with high X-ray luminosity AGN based on their rest-frame hard-band (2-10 keV) X-ray luminosities, instead of the observed full-band (0.5-10 keV) X-ray luminosity as is done in \citetalias{2020MNRAS.497.3273F}.

We calculate the rest-frame hard-band (2-10 keV) X-ray luminosity for our sample of sources crossmatched to the AO13 X-ray data following the methodology in \cite{2019ApJ...876...50L}. We use the \texttt{detml} parameter from the Stripe 82X catalog to determine if a source is significantly detected in a given band-pass. A source is considered significantly detected in the hard or full X-ray band-passes if it has \texttt{detml} $> 15$ in the XMM AO13 catalog \citep{2016ApJ...817..172L}. If a source is significantly detected in the hard-band, we compute the rest-frame hard-band X-ray flux by multiplying the observed hard-band flux by the $k$-correction factor $(1 + z)^{\Gamma - 2}$ and assume a spectral index value of $\Gamma = 1.7$, consistent with a typical AGN spectral slope in the full and hard X-ray bandpasses \citep{2007ApJS..172..341C, 2007ApJS..172..368M, 2006A&A...451..457T, 2004ApJ...600...59K}. If a source is not detected significantly in the hard-band but is detected with a significance above \texttt{detml} $= 15$ in the full-band, we multiply the observed full-band flux by a factor of 0.665 and the same $k$-correction factor as the hard-band, again assuming $\Gamma = 1.7$, to obtain a rest-frame hard-band X-ray flux. If a source is not detected significantly in either the hard or full X-ray band-passes, then we calculate the rest-frame hard-band X-ray flux from the observed soft-band flux by multiplying the soft-band flux by a factor of 1.27. As in \cite{2019ApJ...876...50L}, we assume $\Gamma = 2$ for the soft-band, so no $k$-correction is needed when calculating the rest-frame hard-band X-ray flux from the observed soft-band flux. The factors of 1.27 and 0.665 that we use to convert the observed soft-band and full-band fluxes to the hard-band, respectively, are calculated by assuming an AGN power law spectral model with the spectral index ($\Gamma$) values we use here and extrapolating to estimate the hard-band flux from the soft-band and full-band fluxes. Of the 667 sources we use to perform our analysis, 170 are detected significantly in the hard-band, 466 are detected significantly in the full-band but not the hard-band, and 31 are not detected significantly in either the full or hard band-passes.

We compute the $95 \%$ completeness limit of the rest-frame hard-band X-ray luminosity in each of our redshift ranges of interest using the XMM AO13 flux area curves presented in \cite{2016ApJ...817..172L}. Because the majority ($\sim 70 \%$) of our sample actually have rest-frame hard-band X-ray luminosities computed from the observed full-band flux, we use the observed full X-ray band-pass flux area curves from \cite{2016ApJ...817..172L} to determine a completeness limit for our X-ray sample. We multiply the full band X-ray flux limit by the same conversion factor (0.665) and $k$-correction factor as mentioned above to obtain a rest-frame hard-band X-ray luminosity completeness limit at each redshift range.  The $95 \%$ full-band X-ray flux limit of XMM AO13 is $F_X = 2.09 \times 10^{-14}$ erg s$^{-1}$ cm$^{-2}$. This translates to a $95 \%$ rest-frame hard-band X-ray luminosity of $L_X = 10^{44.08}$ erg s$^{-1}$ at $z=0.75-1.25$, $L_X = 10^{44.41}$ erg s$^{-1}$ at $z=1.25-1.75$, and $L_X = 10^{44.66}$ erg s$^{-1}$ at $z=1.75-2.25$. For the analysis of our sample of galaxies with high X-ray luminosity AGN on the mass-SFR plane, we select only objects having rest-frame hard-band X-ray luminosities above the $95 \%$ X-ray luminosity completeness limit at each corresponding redshift range.

In Table \ref{tab1} we list how many objects are detected in our sample of galaxies with high X-ray luminosity AGN, how many are above the $95 \%$ X-ray luminosity completeness limit, and how many are above both the $95 \%$ X-ray luminosity and stellar mass completeness limits. We note that our X-ray sample selection for the present paper produces a slightly larger sample at our redshift range of interest than the X-ray sample selection used in \citetalias{2020MNRAS.497.3273F}. Using the updated X-ray selection method described above, we find 321 galaxies above the X-ray luminosity completeness limit and 203 galaxies above both the X-ray luminosity and stellar mass completeness limits at $0.75 < z < 2.25$. By contrast, using the X-ray selection method of \citetalias{2020MNRAS.497.3273F}, yields 300 galaxies above the X-ray luminosity completeness limit and 181 galaxies above both the X-ray luminosity and stellar mass completeness limits at $0.75 < z < 2.25$. This means our final mass-complete and X-ray luminosity-complete sample of galaxies with high X-ray luminosity AGN is $\sim 12 \%$ larger when we use the X-ray sample selection detailed in this paper as opposed to that of \citetalias{2020MNRAS.497.3273F}. We note that this updated sample selection process does not require us to rerun the CIGALE SED fitting code on the sample of galaxies with high X-ray luminosity AGN, meaning that the derived stellar masses and SFRs used here are the same as those used in \citetalias{2020MNRAS.497.3273F}.

\section{Demographics of high X-ray luminosity AGN} \label{xray_dems}

The goal of this work is to compare the empirical results of \citetalias{2020MNRAS.497.3273F} to three theoretical simulations. As described in Section \ref{sample_updates}, in order make more direct comparisons to the simulations, we improved upon the sample section of AGN of high X-ray luminosity in \citetalias{2020MNRAS.497.3273F} such that we now select galaxies with high X-ray luminosity AGN based on their hard-band (2-10 keV) X-ray luminosity, rather than their full-band (0.5-10 keV) X-ray luminosity. Before we compare our empirical results to theoretical models, we first perform a consistency check on the distribution and demographics of high X-ray luminosity AGN by measuring the X-ray luminosity function (XLF) of our sample and comparing it to the XLF measured by other studies. We emphasize that the goal of this test is not to focus on the details and implications of the empirical hard-band XLF, but rather to ensure that our sample selection process is sampling the XLF with relative uniformity. We refer the reader to \cite{2019ApJ...871..240A} for the most up-to-date observational constraints of the XLF computed from the cosmic X-ray background (CXB).

\begin{figure*}
\includegraphics[scale=0.7]{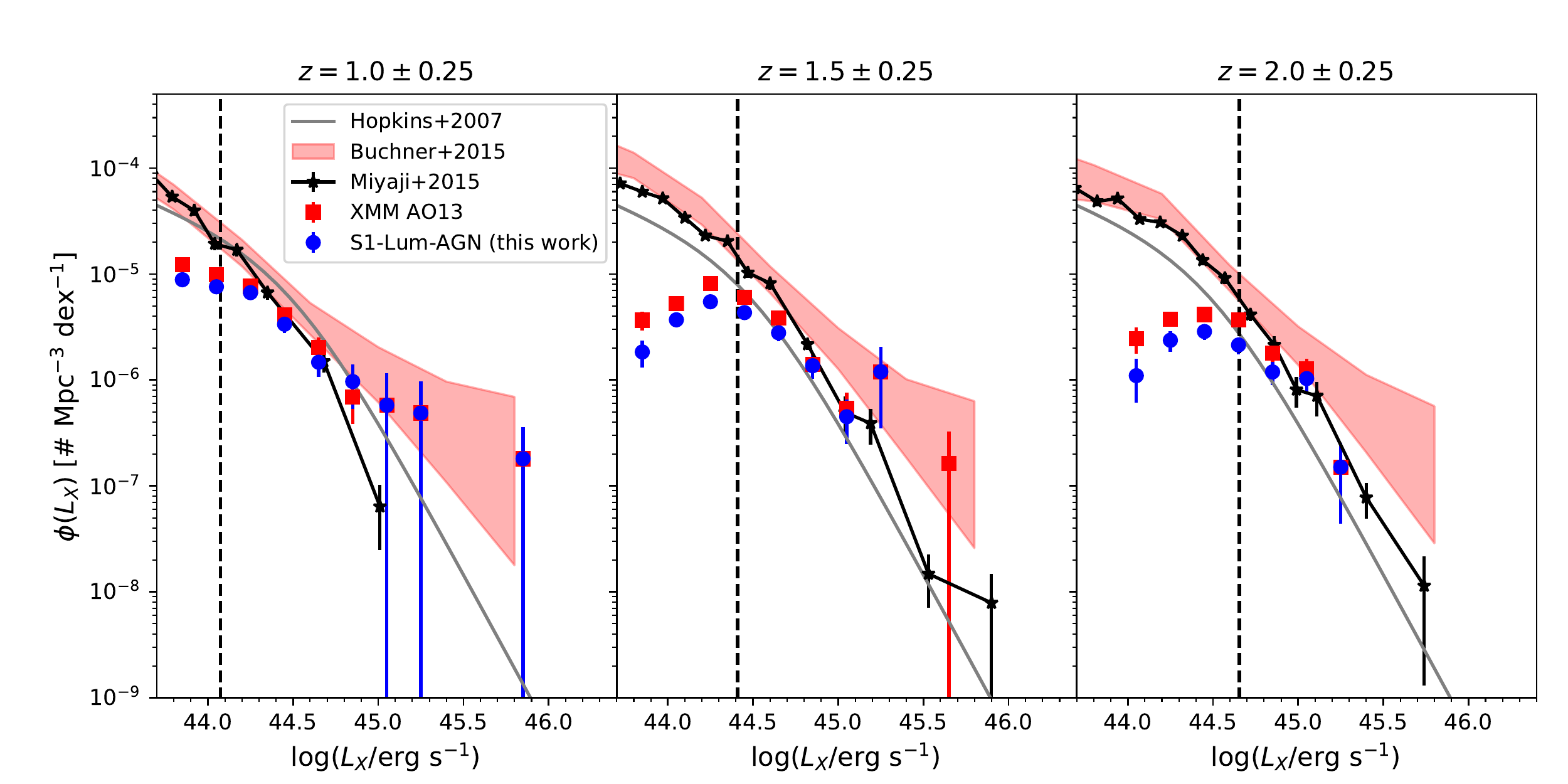}
\caption{We compare the hard (2-10 keV) XLF of our sample of high X-ray luminosity AGN (S1-Lum-AGN, blue), which is based on XMM AO13 sources that have a reliable counterpart in our NEWFIRM $K_S$-selected catalog, to the XLF from all sources in XMM AO13 (red), and to the XLF from published studies \protect\cite{2015ApJ...804..104M} (black, solid), \protect\cite{2015ApJ...802...89B} (red, shaded), and \protect\cite{2007ApJ...654..731H} (gray, solid). The vertical dashed line in each panel indicates the $95 \%$ X-ray luminosity completeness limit, computed from the $95 \%$ flux limit at the upper edge of each redshift bin. The error bars computed for our sample are Poisson errors, and we calculate the XLF using the $1 / V_{\rm max}$ method. We find relatively good agreement between our XLF values and those from the literature, especially at luminosities greater than $L_X = 10^{44.5}$ erg s$^{-1}$.}
\label{obs_XLF}
\end{figure*}

In Figure \ref{obs_XLF}, we plot the observed rest-frame hard-band XLF at three different redshift ranges using our sample of cross-matched galaxies and all galaxies from XMM AO13 that fall in our survey footprint, where SHELA and Stripe 82X overlap, and include the observed XLFs from three other studies \citep{2015ApJ...802...89B, 2015ApJ...804..104M, 2007ApJ...654..731H} for comparison. The \cite{2015ApJ...804..104M} XLF is based on the $Swift$ Burst Alert Telescope (BAT) survey, the \cite{2015ApJ...802...89B} XLF is based on a multi-tiered survey consisting of data from the COSMOS \citep{2007ApJS..172....1S}, CDFS \citep{2011ApJS..195...10X}, AEGIS-XD \citep{2007ApJ...660L...1D}, and XMM-XXL \citep{2016A&A...592A...1P} surveys, while the \cite{2007ApJ...654..731H} XLF is obtained by applying a bolometric correction to the best-fit of the bolometric luminosity function. We include the XLF of the matched and unmatched XMM AO13 sources since the cross-matching can only be done reliably for a subset of the sample \citepalias[see][]{2020MNRAS.497.3273F}. Our XLF is calculated using the $1 / V_{\rm max}$ method \citep[see][]{1968ApJ...151..393S} whereby the luminosity function (LF) of a survey is calculated by dividing the number of galaxies in given bin of luminosity by the bin size and the differential comoving volume which is calculated by taking the difference between the comoving volume ($V_{\rm max}$) at a maximum $z$ that a source of a given luminosity can probe and the comoving volume at the lower edge of the $z$ bin. In all three panels, we show the $95\%$ completeness limit for the corresponding redshift range as the vertical dashed line.

As shown in Figure \ref{obs_XLF}, above the $95 \%$ completeness limit (especially at $L_X \gtrsim 10^{44.5}$ erg s$^{-1}$), our XLF agrees relatively well with the XLF from other studies. For instance, at $z = 0.75-1.25$, our XLF agrees well with the XLF from other studies at $L_X > 10^{44.3}$ erg s$^{-1}$ but at lower X-ray luminosities ($L_X < 10^{44.3}$ erg s$^{-1}$)  there is a small dip that is factor of $\sim 2$ lower for cross-matched sources and a factor of $\sim 1.2-1.5$ lower for the full XMM AO13 XLF. This could be due to uncertainties in our XLF introduced by dust obscuration, causing us to miss AGN at lower $L_X$. At all redshifts, our XLF appears to agree most with the XLF from \cite{2015ApJ...804..104M}, however, they find brighter objects more frequently than we do with X-ray luminosities exceeding $10^{45.5}$ erg s$^{-1}$, which could be due to a number of reasons. Their sample size is much larger than ours, as they use data from the $Swift$ BAT survey which spans an area on-sky of $3.9 \times 10^4$ deg$^2$ and thus they are able to capture much more extreme and rare systems than we can. The $Swift$ BAT survey detects X-rays at energies of $14-195$ keV, however, so \cite{2015ApJ...804..104M} apply a correction to convert X-ray luminosities from $Swift$ BAT to the hard band. This could introduce some source of uncertainty to the measured XLF of \cite{2015ApJ...804..104M} and produce inconsistencies with other studies. We note that other discrepancies between the different empirical XLFs can arise from uncertainties due to corrections for Compton-thick sources, as well as photometric redshift estimates. The latter could especially affect the XLF of \cite{2015ApJ...802...89B} as their sample only has spectroscopic redshifts for $\sim 50 \%$ of sources.

\section{Theoretical Models and Numerical Simulations} \label{simulation_data}
In numerical simulations and theoretical models of galaxy evolution, AGN and SF activity are closely linked and AGN feedback is invoked to regulate galaxy growth. In order to constrain such models, we will explore in Section \ref{results} how three different simulations (i.e., the hydrodynamical simulations IllustrisTNG and SIMBA, and the semi-analytic model SAG) compare to our empirical results on AGN and SF at cosmic noon \citepalias{2020MNRAS.497.3273F}. In this section, we describe the three theoretical models, their implementation of AGN,  stellar, and other feedback, and how we obtain a hard X-ray luminosity from the BHAR in each model. For a recent, comprehensive review of cosmological simulations, we refer the reader to \cite{2020NatRP...2...42V}.

\subsection{IllustrisTNG}
IllustrisTNG is the successor to the Illustris hydrodynamical simulation \citep{2014MNRAS.444.1518V,2014Natur.509..177V, 2014MNRAS.445..175G} and is modeled in varying box sizes and implements improved galaxy physics relative to the original Illustris simulation. The updates to the galaxy physics of Illustris focus on the growth and feedback of massive BHs, galactic winds, and chemical enrichment of gas and stellar evolution \citep{2018MNRAS.473.4077P}. For this project, we utilize the largest box size available, TNG300, which has a volume of $\sim 300^3$ Mpc$^3$ and a mass resolution of $m_{\rm baryon} = 1.1 \times 10^7$ $M_{\odot}$. Following \cite{2020MNRAS.499.4239S}, we utilize the masses and SFRs measured within twice the stellar half mass radius (the $2 \times R_{1/2}$ aperture). The full description of the AGN and stellar feedback implementations in IllustrisTNG, as well as updates to the original Illustris model, can be found in \cite{2017MNRAS.465.3291W} and \cite{2018MNRAS.473.4077P}, respectively. Here, we only give a brief overview. 

One of the ways in which IllustrisTNG improves upon the Illustris model is by replacing the radio-mode AGN feedback in the low-accretion state with a new kinetic AGN feedback model that produces black hole-driven winds. The radio-mode AGN feedback model of Illustris injected thermal bubbles into the ICM, whereas the new kinetic AGN feedback model produces randomly oriented injections of momentum into the surrounding gas. The choice to replace the radio-mode feedback of Illustris is motivated by recent theoretical work that advocates for the inflow/outflow solutions of advection dominated accretion flows (ADAFs) \citep{2014ARA&A..52..529Y}. The new kinetic AGN feedback model is responsible for the quenching of galaxies in massive dark matter halos ($\sim 10^{12} - 10^{14} M_{\odot}$) and for producing `red and dead' populations of galaxies at late times in IllustrisTNG \citep{2017MNRAS.465.3291W}. This in turn produces more realistic SFRs in massive galaxies and estimates of the quenched fraction \citep{2020MNRAS.499.4239S, 2019MNRAS.485.4817D} in IllustrisTNG. In the high-accretion state, both Illustris and IllustrisTNG invoke quasar-mode thermal feedback that heats the gas surrounding the BH. The transition between the low and high-accretion state AGN feedback modes in IllustrisTNG has a dependence on BH mass and Eddington ratio. This choice was made to ensure that BHs can transition to self-regulated states at lower accretion rates, and to ensure that newly seeded BHs do not remain in the low-accretion state.

In addition to AGN feedback, IllustrisTNG also invokes supernova (SN) feedback in order to regulate SF and prevent the over-production of galaxies. The SNe feedback is implemented through wind particles which are injected in random directions, with the strength of the feedback based on the energy released by the SN. In contrast to the original Illustris model, IllustrisTNG injects galactic winds isotropically and modifies the velocity of wind particles by introducing a redshift dependence factor in the calculation of the wind velocity. This choice of adding a redshift-dependent wind velocity is motivated by work done in semi-analytic models that find it necessary to implement a similar approach in order to reproduce the evolution of observed luminosity and stellar mass functions \citep{2013MNRAS.431.3373H}. Additional mechanisms capable of heating and/or removing gas from galaxies, such as tidal stripping, ram-pressure stripping, and dynamical friction all occur naturally in IllustrisTNG. Such mechanisms follow the solutions of the equations of gravity and hydrodynamics in an expanding universe with gravitationally-collapsing structures \citep[see][for review of how these processes affect satellite galaxies in IllustrisTNG]{2019MNRAS.483.1042Y}.

\subsection{SIMBA}
The SIMBA simulation is a hydrodynamical simulation that is run on a box with a side length of 100 Mpc$/h$ and a mass resolution of $m_{\rm gas} = 1.82 \times 10^7$ $M_{\odot}$. SIMBA builds on the MUFASA model \citep{2016MNRAS.462.3265D} by introducing a new black hole growth model based on "torque-limited" accretion \citep{2011MNRAS.415.1027H, 2013ApJ...770....5A, 2015ApJ...800..127A} as well as a new novel sub-grid prescription for AGN feedback. Further details about the SIMBA model can be found in \cite{2019MNRAS.486.2827D}, we briefly summarize some key aspects below.  

The AGN feedback model in SIMBA transfers energy from small to large scales through kinetic outflows with outflow parameters based on observations of AGN feedback. At Eddington ratios greater than 0.02, SIMBA implements a "radiative" feedback mode, modeled after observations of ionised and molecular gas outflows, where the AGN drives multi-phase winds with velocities of $\sim 1000$ km s$^{-1}$. At Eddington ratios lower than 0.2, SIMBA transitions to a "jet" feedback mode where the AGN drives hot gas in collimated jets with velocities on the order of $\sim 10^4$ km s$^{-1}$. Such jets are capable of heating gas on scales up to $\sim 8$ Mpc away from the source. At late times ($z=0$), the jet feedback from numerous AGN is shown to impact the inter-galactic medium (IGM) on large scales in SIMBA \citep{2020MNRAS.499.2617C}. We note that both radiative and jet feedback in SIMBA can simultaneously act on the surrounding gas in systems with Eddington ratios between 0.02 to 0.2. All outflows driven by AGN feedback are purely bipolar, meaning that gas elements are ejected in a direction parallel to the angular momentum vector of the accretion disk. We note that both the radiative and jet feedback modes in SIMBA employ kinetic feedback rather than thermal feedback. In addition to the radiative and jet AGN feedback, SIMBA adds an X-ray heating component to AGN with full velocity jets which heats the gas in galaxies. The X-ray feedback mode is motivated by work done in \cite{2012ApJ...754..125C}, who show that this feedback is capable of quenching massive galaxies in high-resolution zoom simulations.

The stellar feedback model implemented in SIMBA is the same feedback model that is implemented in MUFASA. In MUFASA, stellar feedback is attributed to winds from massive stars as well as long-lived stars (e.g., Type Ia supernovae, asymptotic giant branch stars). The stellar feedback model assumes massive stars drive material out through a combination of Type II supernovae winds, radiation pressure, and stellar winds. This type of kinetic feedback from massive stars is described in detail by \cite{2003MNRAS.339..289S} and \cite{2006MNRAS.373.1265O}. Meanwhile, feedback from long-lived stars is delayed relative to the time of star formation and so is not represented by the same outflows prescribed to massive stars. The prescriptions for Type Ia supernovae feedback and feedback from asymptotic giant branch stars are described in \cite{2005ApJ...629L..85S} and \cite{2015ApJ...803...77C}, respectively. 

\subsection{SAG}
The semi-analytical model SAG is implemented on the MultiDark Planck 2 (MDPL2) dark-matter only simulation with a volume of $1 / h^{3}$ Gpc$^3$. The version of SAG we use in the present paper was also used by \cite{2020MNRAS.499.4239S}. It is run on $9.4 \%$ of the full MDPL2 volume and was provided by the simulation leads (S. Cora, private communication). A full description of the SAG model can be found in \cite{2018MNRAS.479....2C}, here we briefly describe it.  

In SAG, the AGN feedback model is implemented through a radio mode prescription that injects energy into the region surrounding the black hole, thereby reducing hot gas cooling. The accretion of hot gas is assumed to deposit energy into relativistic jets and this energy then gets deposited as heat in the hot gas atmosphere. Black holes in SAG can grow via both radio and quasar mode accretion, however, quasar mode feedback is not implemented in the model. Quasar growth in SAG occurs during mergers and disc instabilities, whereas radio mode accretion arises from the accretion of hot gas onto the BH. Accretion of hot gas onto a BH occurs once a static hot halo has formed around the galaxy hosting the BH and is assumed to accrete continuously with a dependence on BH mass and hot gas mass \citep{2015MNRAS.451.2663H}. For a more detailed description of the AGN feedback model in SAG, we refer the reader to Section 2.2 of \cite{2019MNRAS.483.1686C}. SAG also includes redshift-dependent and virial velocity-dependent stellar feedback that heats gas within the galaxy. The parameter that controls the redshift dependent stellar feedback has been modified to better produce the evolution of the SFR density at high redshifts ($z > 1.5$) \citep{2018MNRAS.479....2C}. In addition, the redshift-dependent stellar feedback implemented here has been shown to produce a quiescent fraction in local galaxies that is in better agreement with observations than other SAMs have been able to achieve \citep{2020MNRAS.498.4327X, 2018MNRAS.479....2C}. We note that \cite{2020MNRAS.499.4239S} compare the empirical quiescent fraction at $1.5 < z < 3$ to SAG and two other semi-analytical models and found that SAG best reproduces the observed trend that quenched fraction increases with stellar mass. This could be attributed to the explicit modeling in SAG of ram-pressure stripping and tidal stripping for satellite galaxies falling into a group or cluster. These processes are not instantaneous, meaning they gradually remove gas from the infalling satellite galaxy.

\subsection{Obtaining an X-ray Luminosity from the BHAR} \label{lx_from_bhar}
In this section, we describe how we obtain a hard-band (2-10 keV) X-ray luminosity from the BHAR in the numerical simulations. By obtaining X-ray luminosities from the predicted BHARs, we can select and analyze samples of highly luminous AGN in the simulations and compare to our observational results of galaxies with high X-ray luminosity AGN. In each simulation, we calculate bolometric luminosities for all objects with BHARs $> 0$ using the following equation:

\begin{equation} \label{lbol}
    L_{\rm bol} = \epsilon \Dot{M} c^2
\end{equation}

\noindent
Here, $\epsilon$ is the radiative efficiency, $\Dot{M}$ is the BHAR, and $c$ is the speed of light. The radiative efficiency $\epsilon$ is set internally by each simulation and is set to 0.1 in SAG and SIMBA, and is set to 0.2 in IllustrisTNG. The radiative efficiency of $\epsilon = 0.2$ in IllustrisTNG was chosen from a range of plausible values, however, lower values would have produced a stronger discrepancy between the measured and fiducial BH accretion rate at high redshift in IllustrisTNG \citep{2017MNRAS.465.3291W}. Once we get a bolometric luminosity we apply two different bolometric corrections (BCs) from \cite{2007ApJ...654..731H} and \cite{2012MNRAS.425..623L} to get hard-band X-ray luminosities.

One of the BCs we apply to $L_{\rm bol}$ to get $L_X$ is that from \cite{2012MNRAS.425..623L} (hereafter \citetalias{2012MNRAS.425..623L}), who use observed BH masses, bolometric luminosities, and X-ray luminosities to derive a relation between the BC and the Eddington ratio $f_{\rm Edd}$ of their AGN samples. In \citetalias{2012MNRAS.425..623L}, they plot $L_{\rm bol} / L_X$ versus $f_{\rm Edd}$ for Type I and Type II AGN and fit a linear relation to the trend. Their Type I sample has higher bolometric luminosities, on average, than their Type II sample. Given that our sample is primarily made up of AGN with high X-ray luminosity, and assuming high X-ray luminosity corresponds to high bolometric luminosity \citep[][]{2020A&A...636A..73D, 2012MNRAS.425..623L, 2007ApJ...654..731H}, we use the best-fit BC given in \citetalias{2012MNRAS.425..623L} for Type I AGN. The BC is obtained using the following equation: 

\begin{equation} \label{L12_BC}
   \log(L_{\rm bol} / L_X) = 0.752 \times \log(\rm{max}[f_{\rm Edd}, 0.0015]) + 2.134
\end{equation}

\noindent
We note that we use $\rm{max}[f_{\rm Edd},0.0015]$ in this equation because values of $f_{\rm Edd}$ lower than 0.0015 can result in an X-ray luminosity that exceeds the bolometric luminosity. The other BC we apply to the simulated galaxies to get $L_{\rm bol}$ is from \cite{2007ApJ...654..731H} (hereafter \citetalias{2007ApJ...654..731H}). In \citetalias{2007ApJ...654..731H}, the authors use fully integrated SEDs of quasars from hard X-rays to radio wavelengths, column densities for a given spectral shape, and X-ray luminosities to derive a relation between X-ray luminosity and bolometric luminosity. The advantage of this method is that the BC only depends on the bolometric luminosity and so it can be applied in reverse to obtain $L_{\rm bol}$ for observed galaxies with high X-ray luminosity AGN. We clarify, however, that we only apply the \citetalias{2007ApJ...654..731H} BC to the simulated galaxies to obtain the predicted X-ray luminosity. The BC for \citetalias{2007ApJ...654..731H} is given by the following equation:

\begin{equation} \label{H07_BC}
    L_{\rm bol} / L_X = 10.83 \left(\frac{L_{\rm{bol}, \odot}}{10^{10} L_{\odot}}\right)^{0.26} + 6.08 \left(\frac{L_{\rm{bol}, \odot}}{10^{10} L_{\odot}}\right)^{-0.02}
\end{equation}

Using these equations (Eqs. \ref{L12_BC} \& \ref{H07_BC}), along with Eq. \ref{lbol}, we convert the BHARs of all sources in the simulations with $\Dot{M} > 0$ to X-ray luminosities in the hard (2-10 keV) band. This allows us to select and analyze samples of galaxies with high X-ray luminosity AGN in the simulations and make direct comparisons to the observed data.

\begin{figure*}
\includegraphics[scale=0.75]{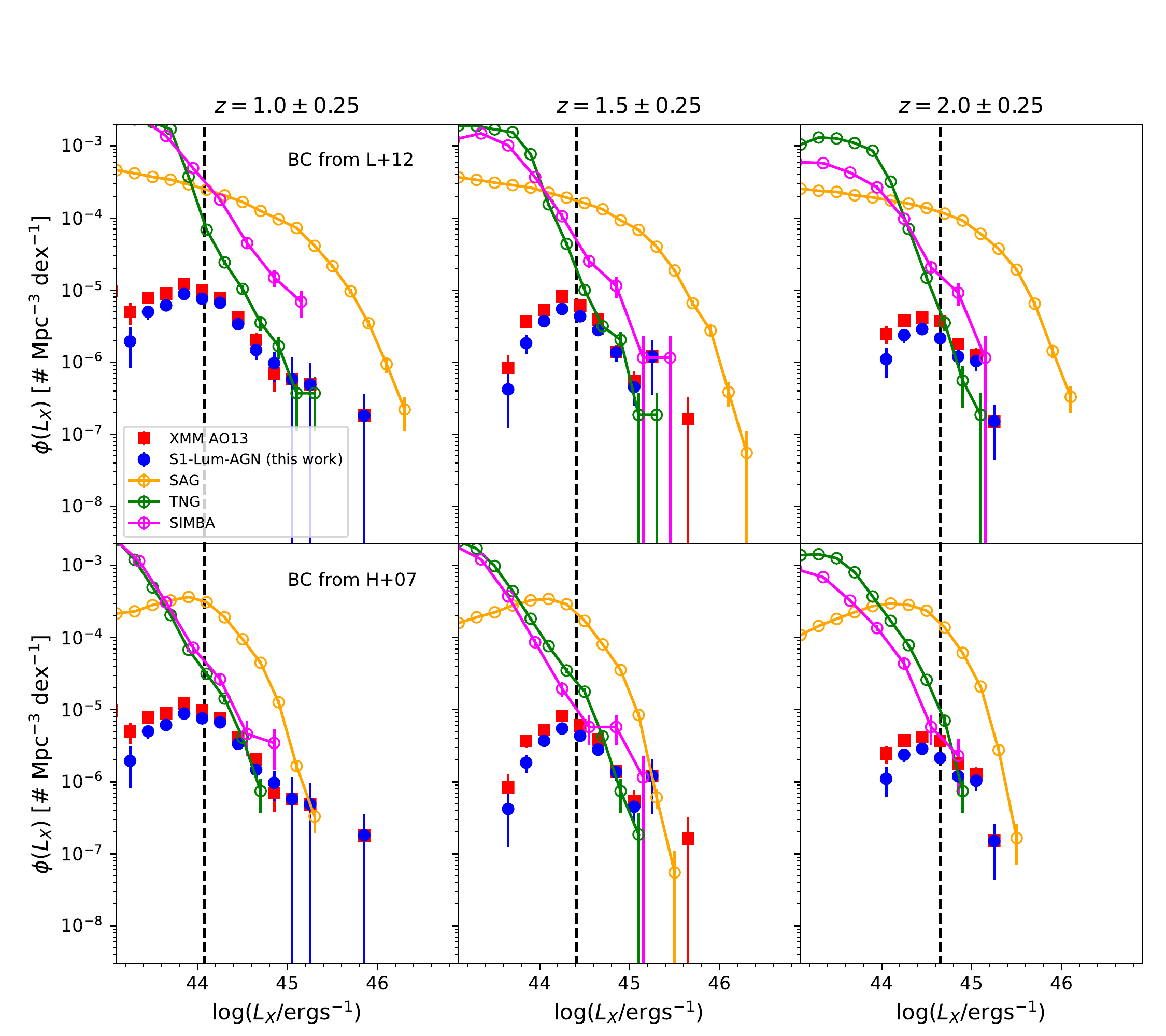}
\caption{The hard (2-10 keV) X-ray luminosity function of our crossmatched sample (blue, same from Figure \ref{obs_XLF}) and the full matched and unmatched XMM AO13 sample that falls in our survey footprint (red, same from Figure \ref{obs_XLF}) compared to the XLF measured for SAG (orange), IllustrisTNG (green) and SIMBA (magenta) in three different redshift bins. All error bars shown are Poisson errors. Each row corresponds to a different bolometric correction that was used to obtain the X-ray luminosity from the bolometric luminosity in the simulations. Row 1 uses the bolometric correction from \protect\citetalias{2012MNRAS.425..623L}, and row 2 uses the bolometric correction from \protect\citetalias{2007ApJ...654..731H}. Regardless of the bolometric correction used, SAG appears to over-predict the number density of galaxies with high X-ray luminosity AGN, relative to the observed sample, at all redshifts and X-ray luminosities. The XLF of IllustrisTNG appears to agree relatively well with the observed XLF, especially when assuming the bolometric correction from \protect\citetalias{2007ApJ...654..731H}. The XLF for SIMBA appears to show that SIMBA over-predicts the number density of galaxies, regardless of bolometric correction used, though not to the extent that SAG does.}
\label{sim_XLF}
\end{figure*}

\section{Comparison of Empirical Results to Theoretical Models} \label{results}
In \citetalias{2020MNRAS.497.3273F}, we explored the distribution of high X-ray luminosity AGN in the SFR-stellar mass plane and found that galaxies with high X-ray luminosity AGN have higher SFR, on average, than galaxies without such AGN at a given stellar mass.  In this paper, we added some refinements to the selection of high X-ray luminosity AGN that increase the fidelity of comparisons to theoretical models and numerical simulations. We now compare our results to the hydrodynamical simulations IllustrisTNG and SIMBA, and the semi-analytic model SAG to explore the following questions: (i) How does the number density of high X-ray luminosity AGN  in theoretical models and numerical simulations compare to the observed distribution (Section \ref{xlf})? (ii) Do theoretical models reproduce the observed distribution of galaxies with and without high X-ray luminosity AGN in the SFR-stellar mass plane (Section \ref{mass_sfr})? (iii) Do they reproduce the empirical result that galaxies with high X-ray luminosity AGN have a higher SFR, on average, at fixed stellar mass than galaxies without such AGN (Section \ref{mass_sfr})?

\subsection{The X-ray Luminosity Function} \label{xlf}
Comparison of our observed XLF (see Figure \ref{obs_XLF}) to the predicted XLF from the simulations can provide important insight to the physical processes directly related to AGN feedback and BH growth in the simulations. In Figure \ref{sim_XLF}, we plot our XLF from the matched and unmatched XMM AO13 source list and the cross-matched sample (same data points from Figure \ref{obs_XLF}) alongside the predicted XLFs from the three different simulations used in this work and the two different BCs, each corresponding to a different row, and at three different redshifts, each corresponding to a different column. The XLF for the simulations is calculated similarly to how it was calculated for the observed sample, but without the $1 / V_{\rm max}$ method, which is not necessary for theoretical models.

Regardless of whether one uses the BC of \citetalias{2012MNRAS.425..623L} or \citetalias{2007ApJ...654..731H}, {\it{SAG appears to over-predict the number density of galaxies with high X-ray luminosity AGN at all redshifts and at almost all but the very highest X-ray luminosities}}, in a few instances, by factors ranging from $\sim 10$ to $\sim 100$. We note that the bright end of the SAG XLF only appears to agree with the data when applying the BC of \citetalias{2007ApJ...654..731H}. When applying the BC of \citetalias{2012MNRAS.425..623L}, SAG appears to produce objects of impossibly high X-ray luminosities ($L_X > 10^{46}$ erg s$^{-1}$ at $z \sim 1.5$ and at $z \sim 2$), which are not seen in our observations and the other observational studies whose XLF we plot in Figure \ref{obs_XLF}. This striking overabundance of high X-ray luminosity AGN in SAG is also evident on the SFR-stellar mass plane (Figure \ref{mass_sfr_scatter}). The discrepancy between the data and SAG is large enough that it cannot be accounted for by the inherent uncertainties associated with computing the XLF for a theoretical simulation using a different BC.

In the case of IllustrisTNG and SIMBA, any disagreement with the data is less pronounced and has a larger dependence on which BC is used to obtain X-ray luminosities in the simulations. For both IllustrisTNG and SIMBA, the best agreement between the empirical and observed XLF is given by the BC of \citetalias{2007ApJ...654..731H}. However, in the highest redshift bin ($z \sim 2$), this BC produces very few objects above the $95 \%$ X-ray luminosity completeness limit and there are no objects above $L_X \sim 10^{45}$ erg s$^{-1}$. By contrast, the BC given by \citetalias{2012MNRAS.425..623L} leads to overproduction of X-ray luminous AGN at $L_X \lesssim 10^{45}$ erg s$^{-1}$ at $z \sim 1$ in SIMBA by a factor of $\sim 10$ relative to the empirical XLF.

It is important to emphasize here the large uncertainties inherent in computing the XLF for simulations of galaxy evolution. One source of uncertainty we have discussed comes from the different BCs that exist in the literature, which all yield different populations of high X-ray luminosity AGN. Another source of uncertainty arises from the calculation of the bolometric luminosity. In this work, we only use a single expression to calculate bolometric luminosity. Some studies, however, suggest that it is necessary to make a distinction between radiatively efficient and inefficient AGN when computing bolometric luminosity \citep{2019A&A...630A..94G, 2012MNRAS.419.2797F, 2005MNRAS.363L..91C}. Radiatively inefficient AGN are typically associated with low accretion rates, relative to the Eddington limit, and low gas densities. For such AGN, the gas flows do not cool efficiently as the radiative cooling is not capable of counteracting the viscosity-generated energy \citep{2014ARA&A..52..529Y}. We consider the effect of using a different expression to calculate bolometric luminosity in the Appendix for the SAG model, as it explicitly separates the radio mode accretion of hot gas from quasar mode accretion, and discuss whether doing so mitigates the overproduction of high X-ray luminosity AGN by SAG (see Figure \ref{A3}). We find that calculating the bolometric luminosity differently for the radio mode accretion hardly affects the XLF predicted by SAG, especially at X-ray luminosities above the observed completeness limits. We note that \cite{2019MNRAS.484.4413H} does this for IllustrisTNG and finds that making a distinction between radiatively efficient and inefficient AGN when computing bolometric luminosity results in an XLF that is largely the same as the XLF that does not make a distinction between radiatively efficient and inefficient AGN, especially at high X-ray luminosities ($L_X > 10^{44}$ erg s$^{-1}$) (see their Figure A1). Lastly, we note that the fraction of Compton-thick AGN is not accounted for in the simulations, making it yet another source of uncertainty. 

\subsection{High X-ray Luminosity AGN on the SFR-Stellar Mass Plane} \label{mass_sfr}
\begin{figure*}
\includegraphics[scale=0.99]{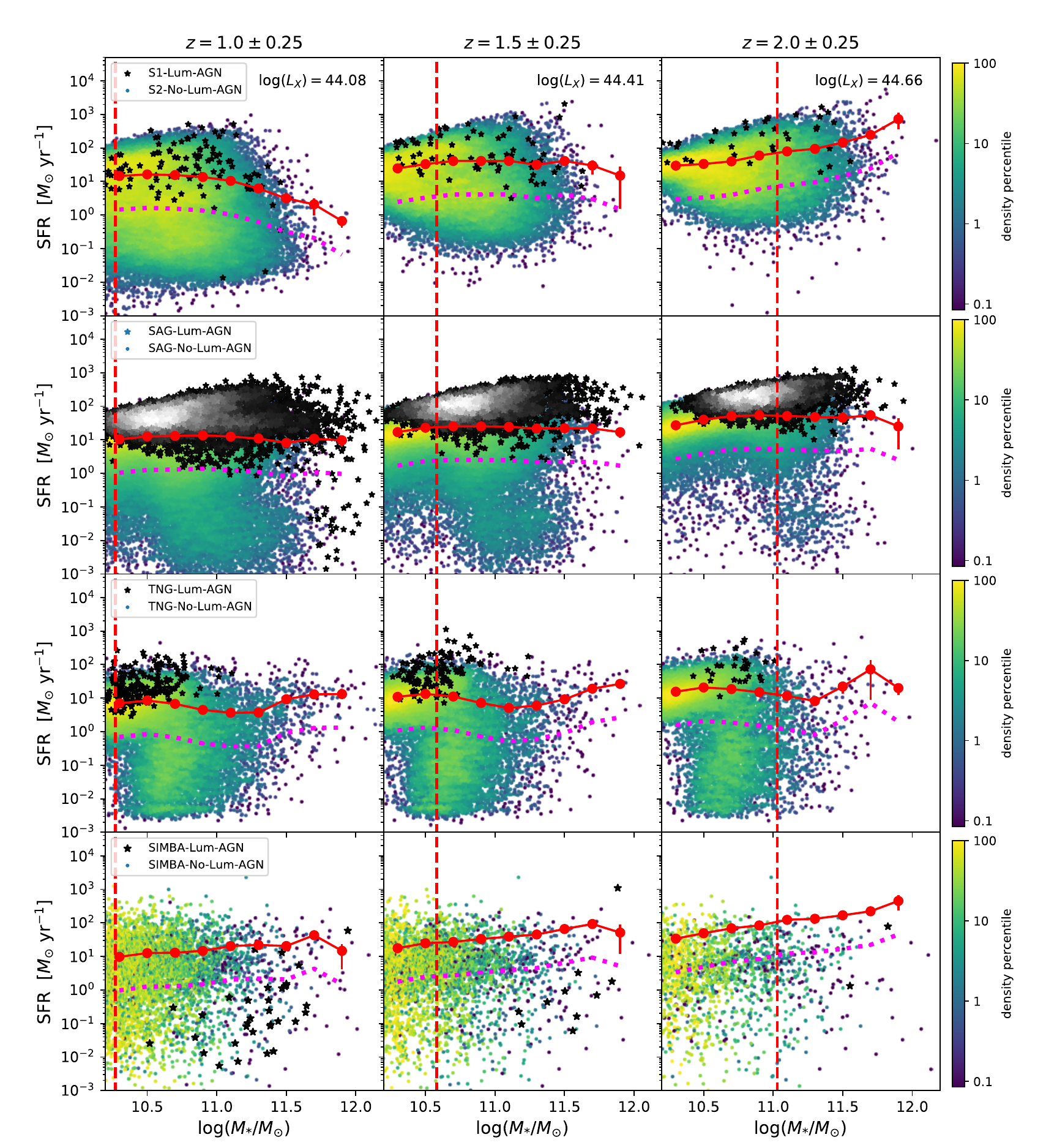}
\caption{The stellar mass-SFR relation for galaxies with (black, stars) and without (colored) high X-ray luminosity AGN for our observed sample (first row), for SAG (second row), for IllustrisTNG (third row), and for SIMBA (fourth row) in three different redshift bins. The galaxies without high X-ray luminosity AGN are color-coded by density on the mass-SFR plane in all panels (see colorbar). The galaxies with high X-ray luminosity AGN in SAG (second row) are similarly color-coded by density on a gray scale to better illustrate their distribution, with black regions having approximately 20 neighbors or less inside an aperture with a diameter of 0.1 dex and white regions having approximately 100 to 180 neighbors within an aperture of the same size. The dashed vertical line shows the observed stellar mass completeness limit at each redshift range. Also shown in each panel is the mean SFR, at fixed stellar mass, for the sample of galaxies without high X-ray luminosity AGN (red circles), which we refer to as the main sequence, and the line that falls 1 dex below the main sequence (dotted magenta). We used the \protect\citetalias{2007ApJ...654..731H} bolometric correction to obtain X-ray luminosities in the models for this figure. Compared to the data, SAG strongly over-produces the number density of AGN of high X-ray luminosity in all three redshift bins, as is seen in Figure \ref{sim_XLF}, while IllustrisTNG shows a lack of high X-ray luminosity AGN at high stellar masses ($M_* > 10^{11} \ M_{\odot}$) at $z \sim 2$. However, both SAG and IllustrisTNG qualitatively reproduce the empirical results that galaxies with high X-ray luminosity AGN have higher mean SFR than galaxies without such AGN at a given stellar mass (see Figure \ref{mean_mass_sfr}). Contrary to the data, in SIMBA the majority of galaxies with high X-ray luminosity AGN appear to have quenched SF.}
\label{mass_sfr_scatter}
\end{figure*}

\begin{figure*}
\includegraphics[scale=0.7]{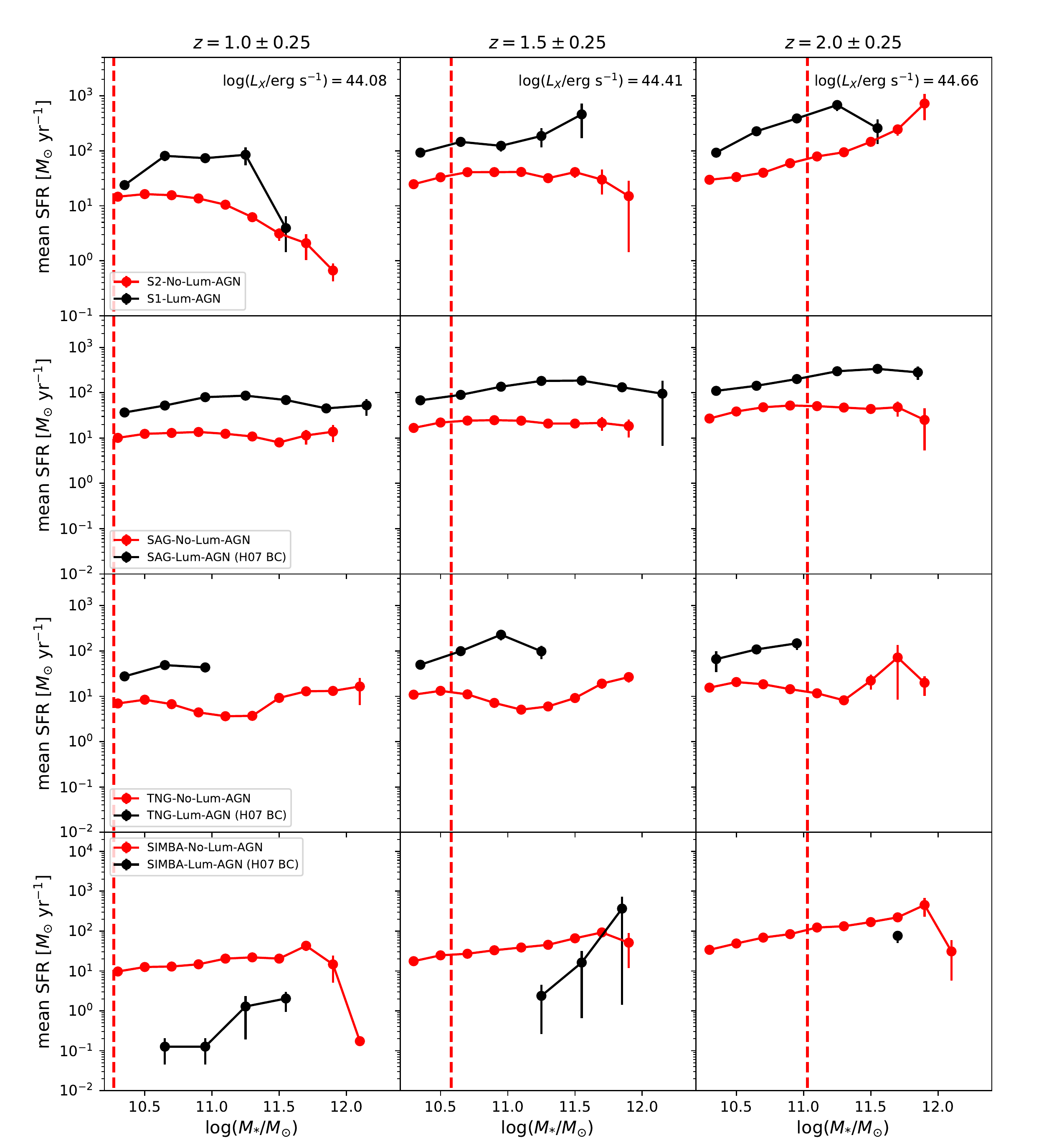}
\caption{Row 1: The mean SFR as a function of stellar mass for the sample of galaxies with (black circles) and without (red circles, same as Figure \ref{mass_sfr_scatter}) high X-ray luminosity AGN. Rows 2-4: Similar to first row, but for SAG (row 2), IllustrisTNG (row 3), and SIMBA (row 4). As in Figure \ref{mass_sfr_scatter}, the BC of \citetalias{2007ApJ...654..731H} was used to obtain hard X-ray luminosities in the models shown here. In qualitative agreement with the data, SAG and IllustrisTNG show enhanced SFR in galaxies with high X-ray luminosity AGN compared to galaxies without such AGN, except at the highest stellar masses in IllustrisTNG. Contrary to the data, galaxies with high X-ray luminosity AGN in SIMBA have a mean SFR, at a given stellar mass, that is lower than the mean SFR of galaxies without such AGN.} 
\label{mean_mass_sfr}
\end{figure*}

In Figures \ref{mass_sfr_scatter} and \ref{mean_mass_sfr}, we show the stellar mass-SFR relation of galaxies with and without high X-ray luminosity AGN for our observed sample and for the three simulations at three different redshift ranges. Figure \ref{mass_sfr_scatter} shows a scatter plot of SFR versus stellar mass, whereas Figure \ref{mean_mass_sfr} shows the mean SFR as a function of stellar mass. Although some refinements were made to the AGN sample selection here based on the hard X-ray luminosity (see Section \ref{sample_updates}), we note that the top rows of Figures \ref{mass_sfr_scatter} and \ref{mean_mass_sfr} essentially represent the empirical results from \citetalias{2020MNRAS.497.3273F} and are similar to Figures 10 and 11 in that paper. We also remind the reader that we use slightly different redshift ranges here so that we can more easily compare the empirical results to the theoretical models.

In Figure \ref{mass_sfr_scatter}, the sample of galaxies without high X-ray luminosity AGN are color-coded by density on the stellar mass-SFR plane, where density is calculated by counting neighbors around each galaxy inside a circle with diameter equal to 0.1 dex. The galaxies with high X-ray luminosity AGN in SAG are similarly color-coded on a gray scale by density to better illustrate their distribution. We also plot the average SFR in different bins of stellar mass for the sample of galaxies without high X-ray luminosity AGN, which we refer to as the main sequence. We calculate $1 \sigma$ errors for the main sequence through a bootstrap method described in \citetalias{2020MNRAS.497.3273F}. This bootstrap method re-samples galaxies $x$ times within a given mass bin, where $x$ is the sample size within that bin, with replacement. For each resampling, we calculate the average SFR and do this 1000 times. By taking the average SFR of all 1000 draws, we generate a set of average SFR values and take the 16th and 84th percentile of this set to calculate the error of the average SFR. Also shown in each panel of Figure \ref{mass_sfr_scatter} is the line that falls 1 dex below the main sequence (magenta, dotted). As is done in \citetalias{2020MNRAS.497.3273F}, we consider any sources that fall below this line to have quenched SFR. In Figure \ref{mean_mass_sfr} we show the mean SFR as a function stellar mass  for galaxies with and without high X-ray luminosity AGN separately. Error bars for all samples are calculated using the aforementioned bootstrap method.

As reported in \citetalias{2020MNRAS.497.3273F}, the top row of Figures \ref{mass_sfr_scatter} and \ref{mean_mass_sfr} show that galaxies with high X-ray luminosity AGN have a higher mean SFR by a factor of $\sim 3-10$ than galaxies without such AGN at a given stellar mass. We suggested in \citetalias{2020MNRAS.497.3273F} that this connection between high SFR and high AGN activity could be partly due to processes where large gas inflows fuel both circumnuclear SF and BH growth \citep[e.g.,][and references therein]{2006LNP...693..143J}. It is also clear from Figure \ref{mass_sfr_scatter} that the vast majority  ($> 95 \%$) of galaxies with high X-ray luminosity AGN do not have quenched SF, thereby suggesting that if AGN feedback quenches SF in a galaxy, it does so after the high X-ray luminosity phase of AGN activity

In Figures \ref{mass_sfr_scatter} and \ref{mean_mass_sfr}, the second, third, and fourth rows from the top show the corresponding distributions of galaxies with and without high X-ray luminosity AGN in the SFR-stellar mass plane in the semi-analytical model SAG (row 2), and the hydrodynamical simulations IllustrisTNG (row 3) and SIMBA (row 4). The results shown in these Figures are based on the BC of \citetalias{2007ApJ...654..731H}. We use the BC prescription of \citetalias{2007ApJ...654..731H} here as it seems to provide the best match between the models and the data (Section \ref{xlf}). These theoretically predicted distributions of galaxies with and without high X-ray luminosity AGN in the SFR-stellar mass plane can be directly compared to the empirical results shown in the top rows of Figures \ref{mass_sfr_scatter} and \ref{mean_mass_sfr}.

We find that SAG strongly over-produces the number density of high X-ray luminosity AGN by factors of $\sim 10-100$. This can be seen in the second row of Figure \ref{mass_sfr_scatter}, as well as in the Figure \ref{sim_XLF} discussed in Section \ref{xlf}. However, SAG does qualitatively reproduce the empirical results showing that galaxies with high X-ray luminosity AGN have higher mean SFR than galaxies without such AGN at a given stellar mass (second row from the top in Figures \ref{mass_sfr_scatter} and \ref{mean_mass_sfr}). As can be seen in Figure \ref{mass_sfr_scatter}, galaxies in SAG with high X-ray luminosity AGN tend to reside in galaxies with enhanced SFR relative to the main sequence and only a very small fraction ($< 1 \%$) of galaxies with high X-ray luminosity AGN in SAG have quenched SFR, consistent with the empirical results.

IllustrisTNG qualitatively reproduce the empirical result that galaxies with high X-ray luminosity AGN have higher mean SFR than galaxies without such AGN activity at a given stellar mass (third row in Figures \ref{mass_sfr_scatter} and \ref{mean_mass_sfr}). As illustrated in Figure \ref{mass_sfr_scatter}, in IllustrisTNG, most galaxies with high X-ray luminosity AGN lie above the main sequence. However, IllustrisTNG shows a lack of high X-ray luminosity AGN at high stellar masses ($M_* > 10^{11} \ M_{\odot}$) at $z\sim 2$ and very few objects with X-ray luminosities above the observed $95 \%$ X-ray luminosity completeness limit at the same redshift.

In the SIMBA simulation, we find at $z < 2$ that galaxies with high X-ray luminosity AGN with $M_* < 10^{11.5} \ M_{\odot}$ have mean SFRs that are 1 to 2 orders of magnitude lower than the main sequence (bottom row of Figure \ref{mean_mass_sfr}). As a result, the mean SFR of galaxies with high X-ray luminosity AGN is lower than the SFR of those without such AGN. This trend is opposite to the empirical results (top row of Figures \ref{mass_sfr_scatter} and \ref{mean_mass_sfr}). Contrary to the data, many high X-ray luminosity AGN in SIMBA have quenched SF, suggesting that AGN feedback or other feedback modes in galaxies with such AGN might be too efficient in SIMBA. We discuss this further in Section \ref{discussion}. In the highest redshift bin, there are very few AGN with $L_X$ above the observational $95 \%$ completeness limit.

We note that the number of galaxies with high X-ray luminosity AGN in the models having quenched SF depends on which BC prescription we use to go from bolometric luminosity to X-ray luminosity (see Section \ref{lx_from_bhar}). Figures \ref{mass_sfr_scatter} and \ref{mean_mass_sfr} currently use the BC from \citetalias{2007ApJ...654..731H}. Figures \ref{A1} and \ref{A2} in the Appendix, show the results with the BC in \citetalias{2012MNRAS.425..623L}. These figures show that the main results remain largely unchanged for all three simulations at $M_* < 10^{11.5} \ M_{\odot}$. We discuss this further in the Appendix.

\section{Discussion} \label{discussion}
In Sections \ref{xlf} and \ref{mass_sfr}, we compare the empirical and predicted XLF and stellar mass-SFR relation of galaxies with and without high X-ray luminosity AGN. These comparisons are critical for guiding how AGN feedback is implemented in theoretical models of galaxy evolution. We discuss the possible sources of discrepancy between the theoretical models and the empirical results, as well as the implications for each theoretical model below.\\

\noindent
\textbf{SAG:} We find that SAG over-produces the number density of high X-ray luminosity AGN at all redshifts and almost all X-ray luminosities by two orders of magnitude. Only at the every highest X-ray luminosities obtained from the \citetalias{2007ApJ...654..731H} BC ($L_X \gtrsim 10^{45}$ erg s$^{-1}$; see Figure \ref{sim_XLF}) does the predicted XLF from SAG appear to match that of data. The overproduction of high X-ray luminosity AGN in galaxies could be due to the lack of quasar mode feedback and/or inefficient stellar feedback allowing high BHARs to go unchecked in such a large fraction of the population. In SAG, the growth of BHs via quasar mode is only triggered by mergers or disc instabilities, meaning such processes would produce high SFRs and AGN activity but may lack the feedback channels to suppress BH growth enough to prevent the overproduction of high X-ray luminosity AGN. The radio mode AGN feedback is only activated through the accretion of hot gas during the cooling process \citep{2018MNRAS.479....2C}, meaning there is no AGN feedback present when BHs grow in the quasar mode (i.e., via mergers or disc instabilities). Observations of quasars have shown that AGN with typically higher accretion rates ($>$ few percent of $f_{\rm Edd}$) are capable of driving molecular and ionised gas outflows with velocities of up to $\sim 1000$ km s$^{-1}$ or more \citep{2017A&A...603A..99P, 2013MNRAS.436.2576L, 2011ApJ...733L..16S}. Therefore, it is not unreasonable to speculate that SAG needs to implement some form of quasar feedback in order to regulate the overproduction of high X-ray luminosity AGN. 

In Figures \ref{mass_sfr_scatter} and \ref{mean_mass_sfr}, we find that an excessive number of high mass galaxies ($M_* > 10^{11.5}$ $M_{\odot}$) in SAG have high X-ray luminosity AGN, but the most massive galaxies in our sample do not. This is consistent with the fact that SAG over-produces the number of high X-ray luminosity AGN at all redshifts. The average SFR of galaxies with and without high X-ray luminosity AGN in SAG follows a trend similar to what is seen in the observations. The mean SFR of galaxies with high X-ray luminosity AGN is higher, by a factor of $\sim 2-10$, than the mean SFR of galaxies without high X-ray luminosity AGN at the same stellar mass. We also find that very few galaxies with high X-ray luminosity AGN have quenched SF, consistent with what we find in the observations. Starbursts in SAG are triggered via mergers and disc instabilities, as is the accretion of cold gas onto the BH in quasar mode. It is therefore no surprise that objects growing their BHs through these mechanisms experience large amounts of SF.\\

\noindent
\textbf{IllustrisTNG:} The XLF of IllustrisTNG agrees with the empirical XLF more than the XLF of the other two simulations, especially when considering the BC of \citetalias{2007ApJ...654..731H} and the BC of \citetalias{2012MNRAS.425..623L} at $z > 1$. We find, however, that even when applying the \citetalias{2007ApJ...654..731H} or \citetalias{2012MNRAS.425..623L} BC to obtain X-ray luminosities, the XLF of IllustrisTNG does not reproduce the bright end of the empirical XLF at $z = 1$ or $z = 2$. This is in agreement with what is found in \cite{2019MNRAS.484.4413H}, who compare the XLF of IllustrisTNG to observations and find that IllustrisTNG cannot produce galaxies with X-ray luminosities at the bright end of the empirical XLF. They suggest that this could be due to the high efficiency of the kinetic mode AGN feedback which suppresses the BH accretion and AGN luminosity. In IllustrisTNG, the kinetic mode AGN feedback is activated at low accretion rates relative to the Eddington limit and has a dependence on BH mass. More specifically, the transition between feedback modes in IllustrisTNG occurs when the Eddington ratio exceeds or drops below a BH mass dependent threshold of $\min [f(M_{\rm BH}),0.1]$, where $f(M_{\rm BH})$ is a quantity proportional to $M_{\rm BH}^2$. As shown in \cite{2017MNRAS.465.3291W}, galaxies with BHs more massive than $\sim 10^8$ $M_{\odot}$ tend to have lower Eddington ratios than galaxies with less massive BHs. The dependence of the kinetic feedback mode on BH mass in IllustrisTNG could mean that galaxies with more massive BHs ($M_{\rm BH} \gtrsim 10^8$ $M_{\odot}$) will always be prevented from producing X-ray luminosities seen at the bright end of the XLF. Galaxies with lower mass BHs ($M_{\rm BH} \leq 10^7$ $M_{\odot}$) cannot produce such luminous AGN either as the bolometric luminosity required to produce X-ray luminosities at the bright end of the observed XLF would exceed the Eddington limit by an order of magnitude or more (e.g., $L_{\rm Edd} = 1.2 \times 10^{45}$ erg s$^{-1}$ for $M_{\rm BH} = 10^7$ $M_{\odot}$). We note that IllustrisTNG explicitly limits BH growth by the Eddington limit by estimating the accretion rate as $\Dot{M}_{\rm BH} =$ min$(\Dot{M}_{\rm Bondi}, \Dot{M}_{\rm Edd})$, where $\Dot{M}_{\rm Bondi}$ is the Bondi accretion rate and $\Dot{M}_{\rm Edd}$ is the Eddington limit \citep{2017MNRAS.465.3291W}. 

In the highest redshift bin of Figures \ref{mass_sfr_scatter} and \ref{mean_mass_sfr}, we do not find any high X-ray luminosity AGN in IllustrisTNG residing in galaxies above our observed stellar mass completeness limit. As discussed above, feedback from AGN with BHs more massive than $10^8$ $M_{\odot}$ is more likely to be kinetic mode feedback. Therefore, if one were to reasonably expect that more massive ($M_* > 10^{11} \ M_{\odot}$) galaxies, with accreting BHs, in IllustrisTNG host more massive BHs ($M_{\rm BH} > 10^8 \ M_{\odot}$), then there is a high probability that the kinetic feedback mode is activated in those galaxies and the feedback, in turn, prevents the accretion rate and the resulting X-ray luminosity from reaching the observed values We note that Figure 6 of \cite{2017MNRAS.465.3291W} explicitly shows how the BHAR drastically drops once the kinetic mode feedback is triggered, which typically occurs at $M_{\rm BH} \gtrsim 10^{8} \ M_{\odot}$. At $z < 2$ we find that the mean SFR, at a given stellar mass, of galaxies with high X-ray luminosity AGN is higher than the mean SFR of galaxies without such AGN in IllustrisTNG (Figures \ref{mass_sfr_scatter} and \ref{mean_mass_sfr}).\\

\noindent
\textbf{SIMBA:} The XLF produced by SIMBA agrees with the empirical XLF to within a factor of $\sim 2-3$ at all redshifts when using the \citetalias{2007ApJ...654..731H} BC to derive X-ray luminosities. However, SIMBA does not seem to be able to produce AGN with X-ray luminosities seen at the bright end of the empirical XLF (with $L_X \gtrsim 10^{45}$ erg s$^{-1}$). A likely reason for this could simply be due to the limited volume of SIMBA ($\sim 3 \times 10^6$ Mpc$^{3}$), as AGN with such high X-ray luminosities are extremely rare and have low number densities ($\lesssim 10^{-7}$ Mpc$^{-3}$). It has been shown in \cite{2021MNRAS.503.3492T} that the radio luminosity function of AGN in SIMBA extends up to the highest observed luminosities, without any evidence for truncation, lending credit to the idea that the lack of AGN with $L_X \gtrsim 10^{45}$ erg s$^{-1}$ in SIMBA is due to limited volume rather than the implementation of BH accretion in the model. We remind the reader that, unlike IllustrisTNG, whose AGN kinetic feedback is injected in randomized directions and is capable of clearing all the gas near the BH, the AGN kinetic feedback in SIMBA is purely bipolar and decoupled from the gas, meaning it explicitly avoids clearing gas in the vicinity of the BH. 

In SIMBA, the mean SFR of galaxies with high X-ray luminosity AGN is lower than the mean SFR of galaxies without high X-ray luminosity AGN. For SIMBA, it is clear that the quenching mechanisms are far too efficient in galaxies with high X-ray luminosity AGN as the SFRs of these objects are far lower than the empirical SFRs of galaxies with high X-ray luminosity AGN. Not only does SIMBA implement jet and radiative mode feedback, it also implements X-ray heating in AGN with full velocity jets. This form of feedback is invoked in gas-poor (i.e., having $M_{\rm gas} / M_* < 0.2$) galaxies when the AGN jet mode feedback is active, meaning two forms of AGN feedback can rapidly heat and eject the gas in galaxies. X-ray feedback is not applied to more gas-rich galaxies under the assumption that such galaxies would be able to absorb and radiate away the X-ray energy \citep{2019MNRAS.486.2827D}. Another possible reason that SF is much lower in galaxies with high X-ray luminosity AGN in SIMBA is because the radiative mode feedback injects momentum, rather than thermal energy (as is done in IllustrisTNG for quasar-mode feedback), into the surrounding gas. Such injections of momentum are more effective at quenching SF than injections of thermal energy as the resulting winds from the kinetic feedback are capable of ejecting gas from the galaxy. It is important to note that due to the small volume of SIMBA, the results we are seeing could also be in part due to environment. The jet mode feedback in SIMBA is shown to have a widespread effect on the IGM \citep{2020MNRAS.499.2617C}, meaning that jet mode feedback could effectively quench the SF of many galaxies existing in large groups or clusters. This would make it such that quasars in groups or clusters have quenched SF, even if the quasar-mode (i.e., radiative) feedback itself is not the primary driver of SF quenching. It is therefore possible that a larger volume could weaken the discrepancy between the empirical results and SIMBA. Future work that explores the characterization of environment in both observations and larger volume simulations will shed light on this issue.

We note that part of the discrepancy we find between SIMBA and the empirical results could be also due to the stochastic accretion of hot gas. The Bondi accretion for a BH could essentially be nonexistent until the BH accretes a single hot gas particle, and then the accretion rate is high for that time step. If this happens to occur during the time step of the output, one gets a very high accretion rate. This was dealt with in \cite{2021MNRAS.503.3492T} by smoothing the accretion rate over a 50 Myr window, which ultimately resulted in lower Bondi accretion rates more consistent with observations. The smoothing of the accretion rate in SIMBA has the potential to prevent quenched galaxies from achieving such high X-ray luminosities and therefore move the mean SFR of galaxies with high X-ray luminosity AGN to higher values and result in slightly better agreement with the empirical results. We note that yet another part of the discrepancy between SIMBA and the empirical results could be due to the fact that the kinetic feedback in SIMBA is bipolar and decoupled from the gas. Since the AGN kinetic feedback avoids clearing gas in the vicinity of the BH, it is possible that the feedback modes cannot prevent BHs in massive quiescent galaxies from accreting too much hot gas and becoming luminous AGN.

\section{Summary} \label{summary}
In \citetalias{2020MNRAS.497.3273F} we explored the relation between AGN and SF activity at cosmic noon. We found that galaxies with high X-ray luminosity AGN have higher mean SFR by a factor of $\sim 3-10$, at a given stellar mass, than galaxies without such AGN. We also found that the vast majority ($> 95 \%$) of galaxies with high X-ray luminosity AGN do not have quenched SF. These results are consistent with a scenario where high SFRs and high AGN luminosities are triggered by mechanisms capable of producing large gas inflow rates into the vicinity of a galaxy's AGN accretion disk. These results also suggest that any quenching from AGN feedback would likely occur after the high X-ray luminosity phase of AGN activity.

In numerical simulations and theoretical models of galaxy evolution, AGN and SF activity are closely intertwined and AGN feedback is routinely invoked to regulate galaxy growth. In this paper, we explored such models by comparing three different simulations (i.e., the hydrodynamical simulations IllustrisTNG and SIMBA, and the semi-analytical model SAG) to our empirical results \citepalias{2020MNRAS.497.3273F} and provide guidance and constraints for future developments and improvement of these models. We made some small refinements to the sample selection in order to increase the fidelity of comparisons to theoretical models. We explored the following questions: (i) How does the number density of high X-ray luminosity AGN in theoretical models and numerical simulations compare to the observed distribution (Section \ref{xlf})? (ii) Do theoretical models reproduce the observed distribution of galaxies with and without high X-ray luminosity AGN in the SFR-stellar mass plane (Section \ref{mass_sfr})? (iii) Do theoretical models reproduce the empirical result that galaxies with high X-ray luminosity AGN have a higher SFR, on average, at fixed stellar mass than galaxies without such AGN (Section \ref{mass_sfr})? We summarize our findings below.

\begin{enumerate}
    \item[(1)] \textbf{SAG:} We find that SAG over-produces the number density of high X-ray luminosity AGN at all redshifts and almost all X-ray luminosities by factors of $\sim 10 - 100$. This is evident in the XLF produced by SAG (Figure \ref{sim_XLF}) and in the distribution of high X-ray luminosity AGN on the SFR-stellar mass plane (Figure \ref{mass_sfr_scatter}). This could be due to a lack of quasar mode feedback and/or inefficient stellar feedback which is allowing high BHARs to go unchecked in a large fraction of the population. Observations of quasars have shown that such AGN are capable of driving outflows with velocities of up to $\sim 1000$ km s$^{-1}$ or more. It is therefore not unreasonable to speculate that SAG may need to implement some form of quasar mode feedback in order to regulate the overproduction of high X-ray luminosity AGN. 

    In Figure \ref{mass_sfr_scatter}, we find an excessive number of high mass galaxies ($M_* > 10^{11.5} M_{\odot}$) in SAG have high X-ray luminosity AGN, consistent with the fact that SAG over-produces the number density of high X-ray luminosity AGN at all redshifts. SAG qualitatively reproduces the empirical result that galaxies with high X-ray luminosity AGN have higher mean SFR, at fixed stellar mass, than galaxies without such AGN by a factor of $\sim 2-10$. SAG also produces very few quenched galaxies with high X-ray luminosity AGN, consistent with what we find in the observations. Starbursts in SAG are triggered via mergers and disc instabilities, so it is no surprise that objects growing their BHs through these mechanisms also experience large amounts of SF.
    
    \item[(2)] \textbf{IllustrisTNG:} The XLF produced by IllustrisTNG shows relatively good agreement with the empirical XLF, especially when considering the BC of \citetalias{2007ApJ...654..731H}. IllustrisTNG, however, cannot seem to produce AGN observed at the bright end of the empirical XLF. This could be due to the high efficiency of the kinetic mode AGN feedback which suppresses BH accretion and therefore AGN luminosity. Because the feedback mode has a dependence on BH mass, it is possible that galaxies with BHs more massive than $\sim 10^8 M_{\odot}$ will always be prevented from producing X-ray luminosities seen at the bright end of the XLF.
    
    When examining the highest redshift bin in Figures \ref{mass_sfr_scatter} and \ref{mean_mass_sfr}, we do not find any high X-ray luminosity AGN residing in galaxies above our observed stellar mass completeness limit in IllustrisTNG. As discussed above and in Section \ref{discussion}, this is likely due to the dependence of the feedback mode on BH mass in IllustrisTNG. The kinetic mode feedback is more likely to be activated in higher stellar mass galaxies ($M_* > 10^{11} \ M_{\odot}$), simply due to higher mass galaxies hosting higher mass BHs, and the feedback in turn prevents the AGN X-ray luminosity from reaching observed values. At $z < 2$, we find the mean SFR, at fixed stellar mass, of galaxies with high X-ray luminosity AGN is higher than the mean SFR of galaxies without such AGN in IllustrisTNG.
    
    \item[(3)] \textbf{SIMBA:} Similar to IllustrisTNG, SIMBA does not seem able to produce AGN with X-ray luminosities seen at the bright end of the empirical XLF (with $L_X \gtrsim 10^{45}$ erg s$^{-1}$). This is could likely be due to the limited volume of SIMBA rather than the BH accretion model itself, as high X-ray luminosity AGN at the bright end of of the XLF are extremely rare and have low number densities. Unlike IllustrisTNG, the bipolar kinetic feedback in SIMBA explicitly avoids clearing out the gas in the vicinity of the BH, meaning that it does not suppress BH growth as effectively as the kinetic AGN feedback model does in IllustrisTNG.
    
    Contrary to the empirical results, the mean SFR of galaxies with high X-ray luminosity AGN is lower than the mean SFR of galaxies without such AGN in SIMBA (Figures \ref{mass_sfr_scatter} and \ref{mean_mass_sfr}). It is clear that the quenching mechanisms in SIMBA act far too efficiently in galaxies with high X-ray luminosity AGN. We note that both forms of AGN feedback in SIMBA (i.e., jet mode and radiative mode) inject momentum into the surrounding gas, rather than thermal energy. Such injections of momentum are capable of removing gas from a galaxy and quenching SF more effectively than injections of thermal energy. In addition to radiative and jet mode feedback, SIMBA implements X-ray feedback in AGN with full-velocity jets meaning multiple forms of feedback can rapidly heat and eject the gas in galaxies. Because the jet mode feedback can affect the IGM on scales up to $\sim 8$ Mpc, we note that the quenching of SF in galaxies with high X-ray luminosity AGN could be partially due to volume and/or environment as a high density environment is much more likely to experience widespread quenching.
    
  We note that part of the discrepancy between SIMBA and the empirical results on the stellar mass-SFR plane could also be due to the stochastic accretion of hot gas onto the BH. If BHARs are smoothed over some range of time, as is done in \cite{2021MNRAS.503.3492T}, it could potentially prevent quiescent galaxies from achieving such high X-ray luminosities and therefore move the mean SFR of galaxies with high X-ray luminosity AGN towards higher values. It is also possible that the bipolar nature of the kinetic feedback model in SIMBA contributes to the disagreement with the empirical results. Since the kinetic feedback is bipolar and decoupled from the gas, the kinetic AGN feedback model in SIMBA may not prevent BHs in massive quiescent galaxies from accreting too much hot gas and becoming luminous AGN.
    
    \end{enumerate}
    
\section*{Data Availability}
The data underlying this work will be shared upon reasonable request to the corresponding author. Email: jflorez06@utexas.edu

\section*{Acknowledgements}

JF, SJ, and YG gratefully acknowledge support from the University of Texas at Austin, as well as NSF grant AST-1413652. JF, SJ, YG, and SLF acknowledge support from NSF grant AST-1614798. JF, SJ, YG, MLS, and SLF acknowledge generous support from The University of Texas at Austin McDonald Observatory and Department of Astronomy Board of Visitors. SLF acknowledges support from the NASA Astrophysics and Data Analysis Program through grants NNX16AN46G and 80NSSC18K0954. MV acknowledges support through NASA ATP grants 16-ATP16-0167, 19-ATP19-0019, 19-ATP19-0020, 19-ATP19-0167, and NSF grants AST-1814053, AST-1814259, AST-1909831 and AST-2007355. SAC acknowledges funding from {\it Consejo Nacional de Investigaciones Cient\'{\i}ficas y T\'ecnicas} (CONICET, PIP-0387), {\it Agencia Nacional de Promoci\'on de la Investigaci\'on, el Desarrollo Tecnol\'ogico y la Innovaci\'on} (Agencia I+D+i, PICT-2018-03743), and {\it Universidad Nacional de La Plata} (G11-150), Argentina. The Institution for Gravitation and the Cosmos is supported by the Eberly College of Science and the Office of the Senior Vice President for Research at the Pennsylvania State University. The authors acknowledge the Texas Advanced Computing Center (TACC) at The University of Texas at Austin for providing High-Performance Computing (HPC) resources that have contributed to the research results reported within this paper. URL: http://www.tacc.utexas.edu. 



\bibliographystyle{mnras}
\bibliography{paper3_agn-sfr} 


\appendix
\section{Applying the L12 Bolometric Correction to Theoretical Models}

\begin{figure*}
\includegraphics[scale=0.99]{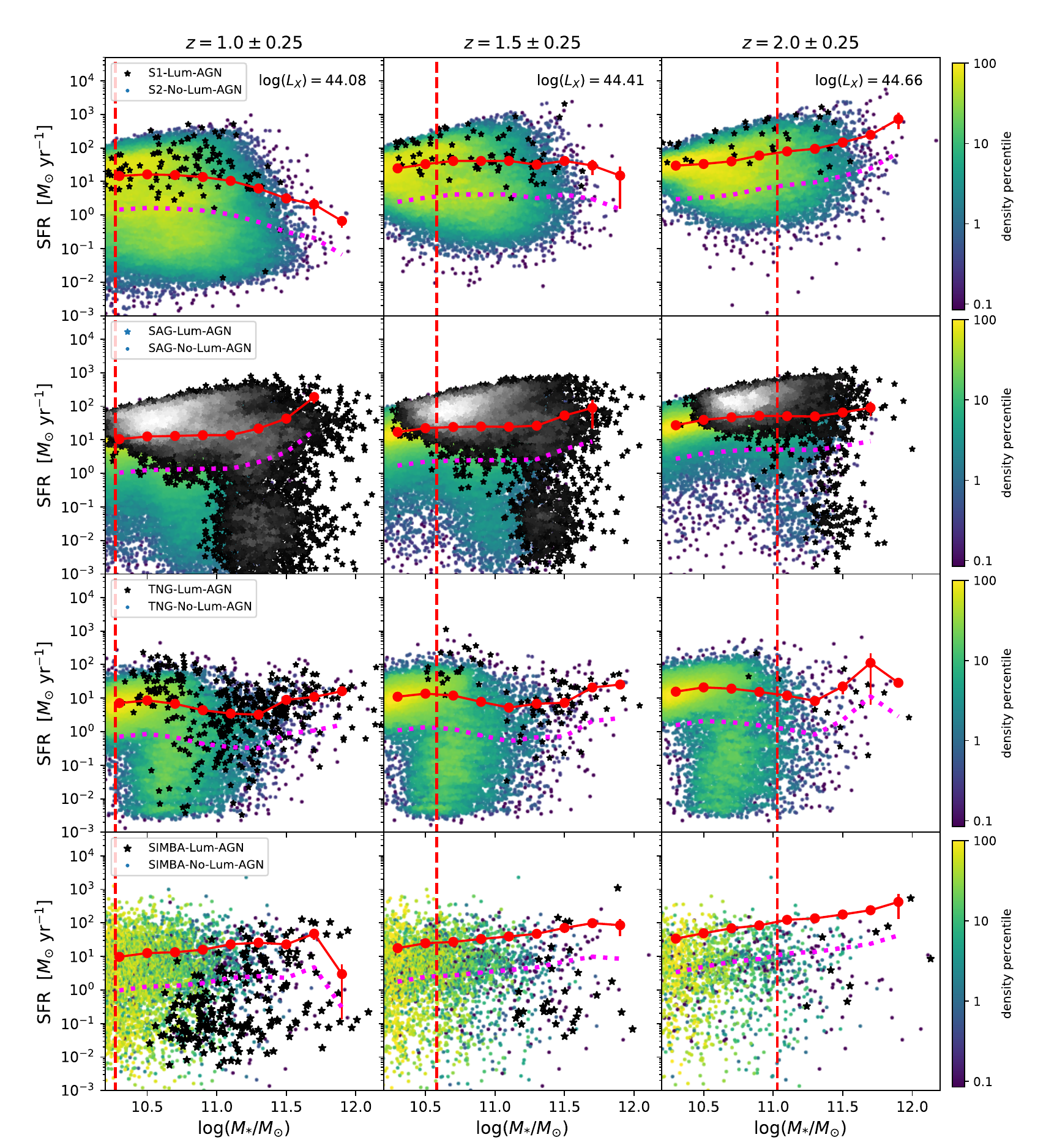}
\caption{The stellar mass-SFR relation for galaxies with (black, stars) and without (colored) high X-ray luminosity AGN for our observed sample (first row), for SAG (second row), for IllustrisTNG (third row), and for SIMBA (fourth row) in three different redshift bins. The galaxies without high X-ray luminosity AGN are color-coded by density on the mass-SFR plane in all panels. As in Figure \ref{mass_sfr_scatter}, the galaxies with high X-ray luminosity AGN in SAG (second row) are color-coded by density on a gray scale to better illustrate their distribution, with black regions having approximately 10 neighbors or less inside an aperture with a diameter of 0.1 dex and white regions having approximately 100 to 150 neighbors within an aperture of the same size. The dashed vertical line shows the observed stellar mass completeness limit at each redshift range. Also shown in each panel is the mean SFR, at fixed stellar mass, for the sample of galaxies without high X-ray luminosity AGN (red circles), which we refer to as the main sequence, and the line that falls 1 dex below the main sequence (dotted magenta). We used the \citetalias{2012MNRAS.425..623L} bolometric correction to obtain X-ray luminosities in the models for this figure. Both SAG and IllustrisTNG appear to predict that the majority of galaxies with high X-ray luminosity AGN have enhanced SFRs, relative to the main sequence, at $M_* \lesssim 10^{11.5} \ M_{\odot}$. The majority of galaxies with high X-ray luminosity AGN in SIMBA, however, appear to have quenched SF. These results are consistent with what we found in Figure \ref{mass_sfr_scatter}}
\label{A1}
\end{figure*}

\begin{figure*}
\includegraphics[scale=0.7]{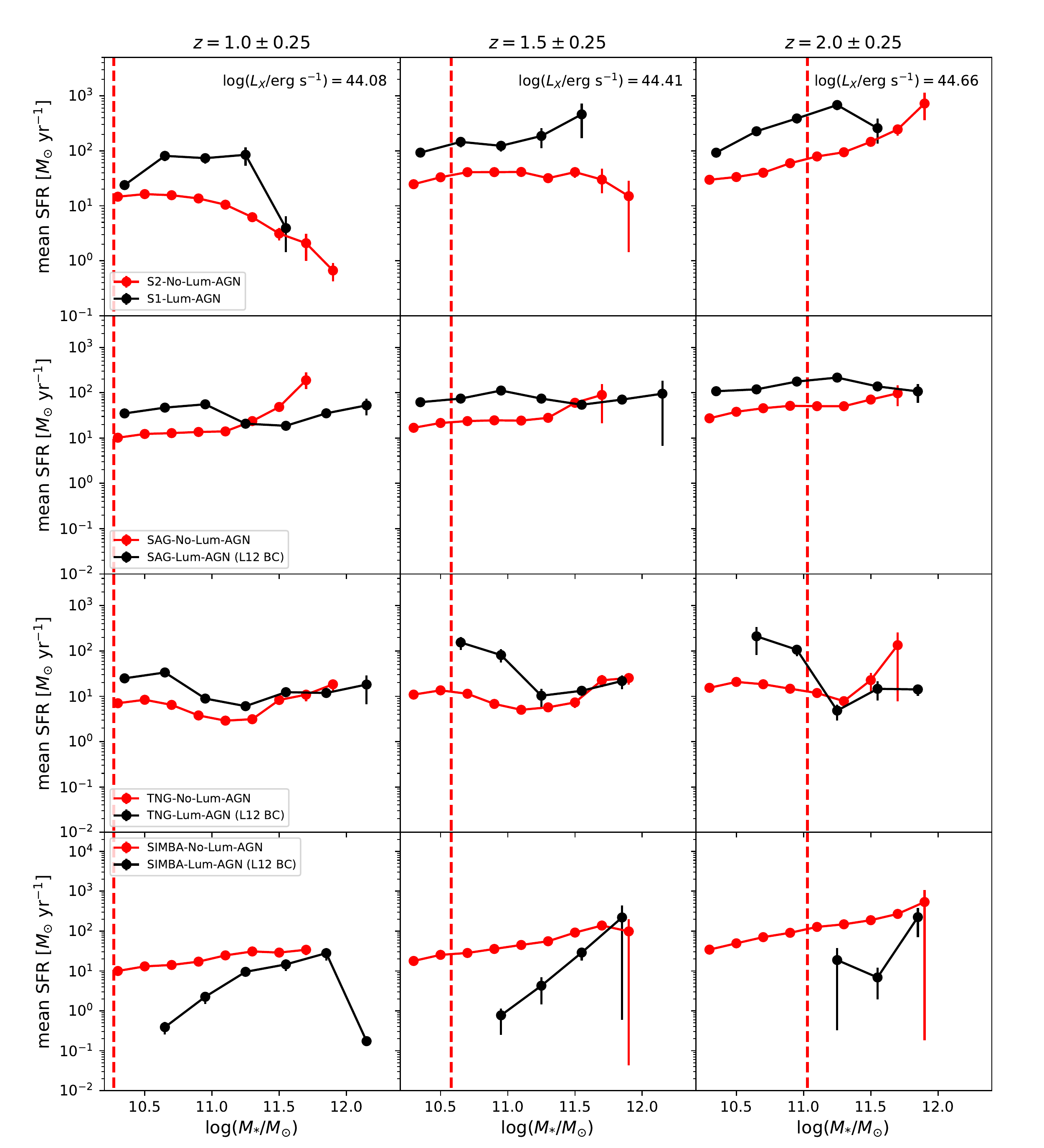}
\caption{The mean SFR as a function of stellar mass for the sample of galaxies with (black circles) and without (red circles) high X-ray luminosity AGN. As in Figure \ref{A1}, the BC of \citetalias{2012MNRAS.425..623L} was used to obtain hard X-ray luminosities in the models shown here. SAG and IllustrisTNG appear to show enhanced SFR in galaxies with high X-ray luminosity AGN compared to galaxies without such luminous AGN, except at the highest stellar mass in IllustrisTNG and SAG ($M_* \gtrsim 10^{11.5} \ M_{\odot}$). SIMBA appears to show that the mean SFR of galaxies with high X-ray luminosity AGN fall below the main sequence, consistent with what we found in Figure \ref{mean_mass_sfr}.}
\label{A2}
\end{figure*}

In this appendix, we follow up on Section \ref{lx_from_bhar} and discuss whether the results change when we derive X-ray luminosities in the theoretical models using the \citetalias{2012MNRAS.425..623L} BC. We plot the same quantities as in Figures \ref{mass_sfr_scatter} and \ref{mean_mass_sfr} using the \citetalias{2012MNRAS.425..623L} BC and show the new results in Figures \ref{A1} and \ref{A2}. The most notable difference in the distributions of high X-ray luminosity AGN we get from the \citetalias{2007ApJ...654..731H} BC to the \citetalias{2012MNRAS.425..623L} BC can be seen in SAG where more of these AGN reside in galaxies with quenched SF and in IllustrisTNG where more high X-ray luminosity AGN reside in more massive galaxies with high and low SF. In Figure \ref{A2}, we see that galaxies with high X-ray luminosity AGN in SAG have lower SFR than galaxies without such AGN at $M* \gtrsim 10^{11.5} \ M_{\odot}$, contrary to what we see in Figure \ref{mean_mass_sfr}. We discuss the changes for each model below when using the \citetalias{2012MNRAS.425..623L} BC. \\

\noindent
\textbf{SAG:} When we apply the \citetalias{2012MNRAS.425..623L} BC to SAG, we find many more galaxies with high X-ray luminosity AGN to have quenched SF, especially in the lowest redshift bin. This causes a turnover in the mean SFR-stellar mass trend where galaxies with high X-ray luminosity AGN go from having higher mean SFR than galaxies without such AGN to having lower mean SFR than galaxies without such AGN at $M_* \gtrsim 10^{11.5} \ M_{\odot}$ at $z \sim 1$. This turnover is not seen in Figure \ref{mean_mass_sfr}, where we use the BC of \citetalias{2007ApJ...654..731H} to obtain X-ray luminosities. We note, however, that the main discrepancies we found between the empirical results and SAG when using the \citetalias{2007ApJ...654..731H} were: (i) The overproduction of high X-ray luminosity AGN (Figures \ref{sim_XLF} and \ref{mass_sfr_scatter}) relative the observed distribution of high X-ray luminosity AGN; (ii) The existence of galaxies with high X-ray luminosity AGN in high mass ($M_* > 10^{11.5} \ M_{\odot}$) galaxies, which is not seen in the observed data. We find that these conclusions remain true even when using the BC of \citetalias{2012MNRAS.425..623L} to obtain X-ray luminosities in SAG. The abundance of quenched galaxies with AGN of high X-ray luminosity arises from the fact that the \citetalias{2012MNRAS.425..623L} BC assigns higher X-ray luminosities to objects with lower Eddington ratios (Equation \ref{L12_BC}). Therefore, many more objects experiencing quenching through radio mode feedback are going to be assigned X-ray luminosities that are above the observed $95 \%$ X-ray luminosity completeness limit when we apply the \citetalias{2012MNRAS.425..623L} prescription to obtain X-ray luminosity. Lastly, we note that only $\sim 10 \%$ of the high X-ray luminosity AGN actually reside in galaxies with quenched SF when applying the BC of \citetalias{2012MNRAS.425..623L} to galaxies in SAG.   \\

\noindent
\textbf{IllustrisTNG:} It is notable that when using the \citetalias{2012MNRAS.425..623L} BC to obtain X-ray luminosities for AGN in IllustrisTNG, there are high X-ray luminosity AGN present in galaxies with $M_* > 10^{11}$, given that many of these objects are missing in row 3 of Figures \ref{mass_sfr_scatter} and \ref{mean_mass_sfr}. We find that at higher stellar masses, the mean SFR of galaxies with high X-ray luminosity AGN matches the sample of galaxies without high X-ray luminosity AGN in IllustrisTNG (Figure \ref{A2}). At higher stellar masses, galaxies with high X-ray luminosity AGN have BHs accreting at lower rates relative to the Eddington limit, meaning the efficient AGN kinetic feedback mode is activated in these systems. As mentioned in Section \ref{lx_from_bhar}, the BC of \citetalias{2012MNRAS.425..623L} assigns higher X-ray luminosities to objects with lower Eddington ratios. This is why the BC of \citetalias{2012MNRAS.425..623L} picks up more galaxies with high X-ray luminosity AGN above the $95 \%$ X-ray luminosity completeness limit at $M_* > 10^{11}$ $M_{\odot}$ in all three simulations than the \citetalias{2007ApJ...654..731H} BC. What the \citetalias{2012MNRAS.425..623L} BC is doing is essentially producing higher X-ray luminosities for objects with low Eddington ratios ($f_{\rm Edd} \lesssim 0.1$). Such AGN tend to exist in higher mass galaxies and are accreting in the kinetic mode in the case of IllustrisTNG. Such galaxies will have SF that is quenched by the kinetic mode feedback of IllustrisTNG. \\

\noindent
\textbf{SIMBA:} We find very little change, qualitatively, between the results given by the \citetalias{2012MNRAS.425..623L} BC and \citetalias{2007ApJ...654..731H} BC for SIMBA. The results given by the \citetalias{2012MNRAS.425..623L} BC remain consistent to what we found previously for the \citetalias{2007ApJ...654..731H} BC, with the only real notable difference being seen at the lowest and highest redshift bins where more high X-ray luminosity AGN are produced by the \citetalias{2012MNRAS.425..623L} BC.

\section{Calculating Bolometric Luminosity with a Dependence on Eddington Ratio}
In this appendix, we follow up on Section \ref{lx_from_bhar} and discuss the effect of making a distinction between radiatively efficient and inefficient AGN when computing bolometric luminosity. We consider using a different expression that depends on the Eddington ratio to calculate bolometric luminosity for the SAG model, as it explicitly separates the radio mode accretion of hot gas from quasar mode accretion.

\begin{figure*}
\includegraphics[scale=0.75]{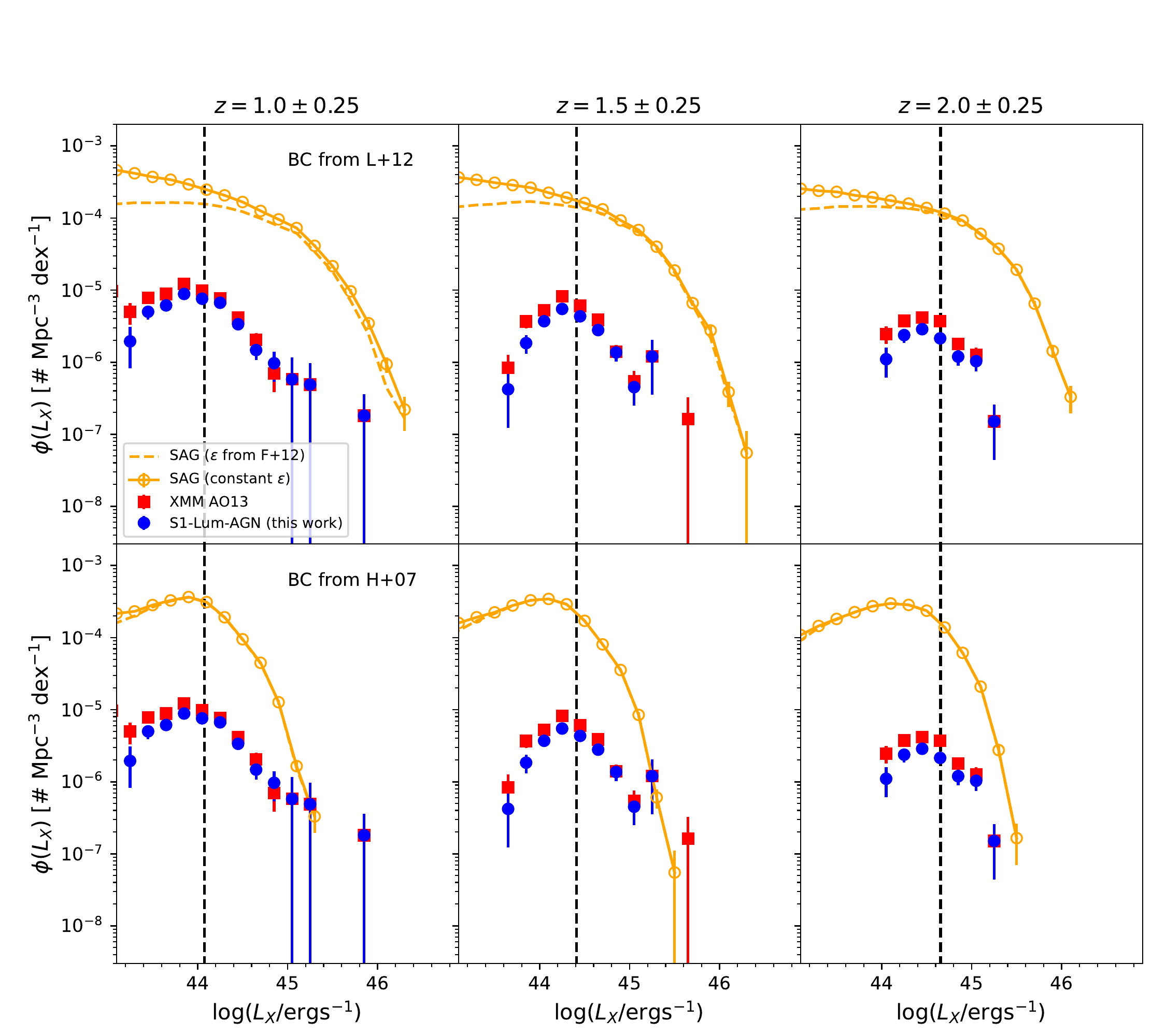}
\caption{Similar to Figure \ref{sim_XLF} which shows the XLF of our observed sample (blue), XMM AO13 (red), and the theoretical models using the \protect\citetalias{2012MNRAS.425..623L} BC (top row) and \protect\citetalias{2007ApJ...654..731H} (bottom row) BC. We remind the reader that the dashed vertical line represents the X-ray luminosity completeness limit. Here, we only focus on the SAG model (orange) and show the predicted XLF that arises when we apply two different methods to obtain bolometric luminosity. The SAG XLF with the error bars and circles is the same that is shown in Figure \ref{sim_XLF}, where the radiative efficiency $\epsilon$ is treated as a constant. The SAG XLF shown by the dashed line is computed by treating the radiative efficiency of the radio accretion mode as a variable that depends on the Eddington ratio \protect\citep[from][]{2012MNRAS.419.2797F}. The only difference between the two predicted XLFs produced by SAG is seen at the faint end of the XLF in the top panels, where the BC of \protect\citetalias{2012MNRAS.425..623L} is applied to get X-ray luminosity. The faint end slope of the SAG XLF predicted by the variable radiative efficiency slightly flattens out relative to the SAG XLF predicted by assuming a constant radiative efficiency. We find that SAG still overproduces high X-ray luminosity AGN and the XLF remains largely unchanged when assuming a variable radiative efficiency to calculate bolometric luminosity.}
\label{A3}
\end{figure*}

For sources accreting in the radio mode in SAG, we compute their bolometric luminosities using Eq. (7) of \cite{2012MNRAS.419.2797F} and recalculate their X-ray luminosities using the BCs of \citetalias{2007ApJ...654..731H} and \citetalias{2012MNRAS.425..623L}. We note that all sources accreting in radio mode in SAG have very low Eddington ratios ($f_{\rm Edd} < 0.01$). We show the predicted XLF by SAG in Figure \ref{A3} when assuming a constant radiative efficiency for all sources (orange circles with error bars, same as in Figure \ref{sim_XLF}), and when using Eq. (7) of \cite{2012MNRAS.419.2797F} to calculate bolometric luminosity for AGN accreting in radio mode (orange, dashed line). Again, we show the XLFs when applying the \citetalias{2012MNRAS.425..623L} BC to the model (top panels), as well as the \citetalias{2007ApJ...654..731H} BC (bottom panels). The only difference between the predicted XLFs produced by SAG can be seen in the top panels, where the BC of \protect\citetalias{2012MNRAS.425..623L} is applied to get X-ray luminosity. The faint end slope of the SAG XLF predicted by the variable radiative efficiency slightly flattens out relative to the SAG XLF predicted by assuming a constant radiative efficiency, however, both XLFs predicted by SAG in the top and bottom panels are largely the same in shape and normalization above the observed X-ray luminosity completeness limit. We find that SAG still overproduces high X-ray luminosity AGN and the XLF remains largely unchanged when assuming a variable radiative efficiency to calculate bolometric luminosity.

\bsp	
\label{lastpage}
\end{document}